\documentclass[10 pt, onecolumn, aps, prd, amsmath, superscriptaddress, amssymb, floatfix]{revtex4}
\usepackage[cp1251]{inputenc}
\usepackage{mathtext}
\usepackage{color}
\usepackage[T2A]{fontenc}
\usepackage[paperwidth=210mm,paperheight=297mm,top=1.5cm,bottom=1.5cm,left=2.5cm,right=1cm]{geometry}
\DeclareSymbolFont{T2Aletters}{T2A}{cmr}{m}{rm}
\usepackage[russian, english]{babel}

\usepackage{epsfig}
\usepackage{epstopdf}

\begin{document}
\renewcommand{\figurename}{Fig.}
\title{Decays of a neutral particle with zero spin and arbitrary \textbf{\textit{CP}} parity into two off-mass-shell \textbf{\textit{Z}} bosons}
\author{T.V.~Zagoskin} \email{taras.zagoskin@gmail.com}
\affiliation{NSC ``Kharkov Institute of Physics and Technology'', 61108 Kharkov, Ukraine}
\author{A.Yu.~Korchin} \email{korchin@kipt.kharkov.ua}
\affiliation{NSC ``Kharkov Institute of Physics and Technology'', 61108 Kharkov, Ukraine}
\affiliation{V.N.~Karazin Kharkov National University, 61022 Kharkov, Ukraine}

\begin{abstract}
Effects are investigated of $CP$ symmetry violation in the decay of a scalar particle $X$ (the Higgs boson) into two off-mass-shell $Z$ bosons both decaying into a fermion-antifermion pair, $X \rightarrow Z_1^* \, Z_2^* \rightarrow f_1 \bar{f}_1 \, f_2 \bar{f}_2$. The most general form of the amplitude of the transition $X \rightarrow Z_1^* Z_2^*$, wherein the boson $X$ may not have definite $CP$ parity, is considered. Limits of applicability of the narrow-$Z$-width approximation used when obtaining differential widths of the decay under consideration are determined. Various observables connected with the structure of the amplitude of the decay $X \rightarrow Z_1^* Z_2^*$ are studied. These observables are analyzed in the Standard Model, as well as in models conceding indefinite $CP$ parity of the Higgs boson. An experimental measurement at the LHC of angular and invariant mass distributions of the decay $X \rightarrow Z_1^* Z_2^* \rightarrow f_1 \bar{f}_1 f_2 \bar{f}_2$ can give information about the $CP$ properties of the Higgs boson and its interaction with the $Z$ boson.
\end{abstract}

\maketitle

\section{Introduction}

In 2012 the ATLAS and CMS collaborations detected \cite{A discovery of a boson which looked like the SM Higgs boson} a neutral boson $h$ with a mass of about 126 GeV. At the present time, detailed study of properties of this particle, called the Higgs boson, is an important task. The Standard Model (SM) Higgs boson is a state with $J^{CP}=0^{++}$, and all the available experimental data about properties of the particle $h$ are close to the corresponding theoretical predictions about the SM Higgs boson (see, for example, \cite{h->Z_1^* Z_2^*->4l and h->W-* W+*->l nu l nu ATLAS, h->Z_1^* Z_2^*->4l CMS, the probability that CPh=-1 is very small CMS}). In particular, the spin of the boson $h$ is equal to zero or two, and many hypotheses in which the spin of $h$ is two are excluded with probability 95\% or higher \cite{h->Z_1^* Z_2^*->4l CMS}. At the same time, the situation may be more complicated. For example, some supersymmetric models predict existence of neutral bosons with negative or even indefinite $CP$ parity \cite{Pilaftsis:1999np,Barger:2009pr,Branco:2012pre}.

The issue of the $CP$ parity of the Higgs boson is also related to the search for $CP$ symmetry breaking sources which are additional to the mechanism built into the Cabibbo-Kobayashi-Maskawa quark-mixing matrix. Such sources of $CP$ violation, for example, in the Higgs sector, could help in explaining the known problem of the matter-antimatter asymmetry in the Universe \cite{Bailin}.

It has been suggested \cite{Voloshin:2012,Bishara:2013vya} that the $CP$ properties of the Higgs boson be studied by investigation of decays into two photons, $h \to \gamma \gamma$, via measurement of the polarization characteristics of the photons. In Refs. \cite{Korchin:2013} the decay to the photon and the $Z$ boson, $h \to Z^* \gamma \to f \bar{f} \gamma$, has been examined while \cite{Gainer:2011aa,Korchin:2014kha} study the decay to the photon and a lepton pair, $h \to \gamma l^+ l^-$. In these papers it has been shown that the ``forward-backward'' escape asymmetry for the final fermions carries information about the $CP$ properties of the $h$ boson and physics beyond the SM.

Investigation into the decay of the Higgs boson into two $Z$ bosons with their consequent decay to fermions is another opportunity to ascertain the $CP$ properties of $h$. Such a cascade decay wherein the final fermions are leptons, along with the two-photon decay channel, has allowed the determination \cite{A discovery of a boson which looked like the SM Higgs boson} of the mass of the particle $h$ with the highest accuracy. In Refs. \cite{amplitudes of X->ZZ and X->W-W+, distributions of a_2 Choi, distributions of a_2 Menon, distributions of a_2 Sun} theoretical distributions of the decay $h \to Z_1^* Z_2^* \to f_1 \bar{f}_1 f_2 \bar{f}_2$ have been studied at various values of the spin of $h$ and in case of various $CP$ properties of this boson. In \cite{distributions of a_2 Choi} it has been reported what properties of experimental distributions testify about a particular spin and a particular $CP$ parity of $h$. In \cite{amplitudes of X->ZZ and X->W-W+, distributions of a_2 Menon, distributions of a_2 Sun} asymmetries measurement of which allows clarification of the mentioned properties of the Higgs boson are suggested and investigated. Finally, papers \cite{methodologies on constraining the Higgs boson couplings to ZZ W-W+ gamma gamma and Z gamma from experimental data} put forward various methodologies on getting constraints on the Higgs boson couplings to $ZZ$, $W^- W^+$, $\gamma \gamma$ and $Z \gamma$ from experimental data.

Besides, various theories with spontaneous breaking of the conformal invariance (for example, theories of technicolor) assume the existence of one more neutral zero-spin particle which interacts with the gauge bosons -- the dilaton. At present, the mass of the dilaton is not determined, but according to estimates performed in Ref. \cite{the dilaton mass}, in some models the mass can exceed $10^4$ GeV. Along with that, in \cite{Is h the SM Higgs boson or the dilaton? number1, Is h the SM Higgs boson or the dilaton? number2, Is h the SM Higgs boson or the dilaton? (a continuation of number2)} it has been shown that the variant in which the boson $h$ is the dilaton is not excluded.

In order to clarify the $CP$ properties of the particle $h$ and the hypothetical dilaton we consider a neutral particle $X$ with zero spin and arbitrary $CP$ parity. We examine the decay $X \rightarrow Z_1^* Z_2^* \rightarrow f_1 \bar{f}_1 f_2 \bar{f}_2$ in case of the non-identical fermions, $f_1 \neq f_2$, and study in detail the differential width of this decay with respect to the three angles of the fermions in the helicity frame and with respect to the invariant masses of the fermion pairs $f_1 \bar{f}_1$ and $f_2 \bar{f}_2$. The most general $X \to Z_1^* Z_2^*$ vertex, which generalizes the corresponding SM vertex and contains a term corresponding to the negative $CP$ parity of the particle $X$, is used.

We also find limits of applicability of the narrow-width approximation for the $Z$ boson for the presented calculation of differential widths of the given decay. By means of this approximation we derive a formula for the total width of the decay $X \rightarrow Z_1^* Z_2^* \rightarrow f_1 \bar{f}_1 f_2 \bar{f}_2$ (the formula is valid also in case $f_1 = f_2$) and a formula for the total width of the decay $h \rightarrow Z_1^* Z_2^*$. These formulas are more general and more precise than those obtained in Ref. \cite{Gamma (tilde Gamma) in the SM calculated by means of the narrow-Z(W)-width approximation}.

Next we find observables connected with the structure of the amplitude of the decay $X \rightarrow Z_1^* Z_2^*$. The formula for the fully differential decay width contains nine coefficients related to the amplitude $X \rightarrow Z_1^* Z_2^*$. For each of them one or two observables linear in this coefficient are defined. Note that some of these observables, as well as different ones, have been studied in  \cite{amplitudes of X->ZZ and X->W-W+, Observables connected with the CP properties of the Higgs boson and defined with the momenta of the fermions, distributions of a_2 Menon, distributions of a_2 Sun}, however we also obtain new experimentally measurable quantities and analyze the dependences of the observables on the mass of one of the $Z$ bosons ($Z_2^*$) in much more detail than it has been done in the mentioned papers. This analysis is carried out within the framework of the SM as well as in certain SM extensions wherein the boson $h$ is a mixture of a $CP$-even state and a $CP$-odd one. Measurement of the suggested observables at the LHC can yield important information about the $CP$ properties of the Higgs boson and its interaction with the $Z$ boson.


\section{Formalism for the decays $X \rightarrow Z_1^* Z_2^* \rightarrow f_1 \bar{f}_1 f_2 \bar{f}_2$}
\label{a study of decays X-> Z_1^* Z_2^* -> f_1 antif_1 f_2 antif_2}

\subsection{The amplitude of the decay $X \rightarrow Z_1^* Z_2^*$ and the fully differential decay width for $X \rightarrow Z_1^* Z_2^* \rightarrow f_1 \bar{f}_1 f_2 \bar{f}_2$}

Let us consider the decay of a neutral spin-zero particle $X$ with arbitrary $CP$ parity into two off-mass-shell $Z$ bosons ($Z_1^*$ and $Z_2^*$) each of which decays to a fermion-antifermion pair, $f_1 \bar{f}_1$ and $f_2 \bar{f}_2$,
\begin{align}
\label{X-> Z_1^* Z_2^* -> f_1 antif_1 f_2 antif_2}
X \rightarrow Z_1^* Z_2^* \rightarrow f_1 \bar{f}_1 f_2 \bar{f}_2,
\end{align}
where $m_X > 2 (m_{f_1} + m_{f_2})$ (to satisfy the law of conservation of energy in a rest frame of $X$), $m_X$ is the mass of the particle $X$, $m_{f_j}$ is the mass of the fermion $f_j$. We will consider this decay at tree level. If $m_X \in (4 m_b, 2 m_t]$ ($m_b$ is the mass of the $b$ quark, $m_t$ is the mass of the $t$ quark), which holds true if $X = h$, then $f_j = e^-, \mu^-, \tau^-, \nu_e, \nu_{\mu}, \nu_{\tau}, u, c, d, s, b$. If $m_X > 4 m_t$, which is possible \cite{the dilaton mass} if $X$ is the dilaton, then $f_j$ can be the top quark as well.

From the energy-momentum conservation we find that $a_1$ and $a_2$ ($a_j$ is the mass squared of the boson $Z_j^*$, i.e. the invariant mass squared of the pair $f_j \bar{f}_j$) lie within limits
\begin{equation}
\label{intervals of a_1 and a_2}
4 m_{f_1}^2 < a_1 <  (m_X - 2 m_{f_2})^2, \qquad 4 m_{f_2}^2 < a_2 < (m_X - \sqrt{a_1})^2.
\end{equation}

The amplitude $A_{X \rightarrow  Z_1^* Z_2^*} (\lambda_1, \lambda_2)$ of the decay of $X$ into $Z_1^*$ and $Z_2^*$ is equal to \cite{amplitudes of X->ZZ and X->W-W+, distributions of a_2 Menon, distributions of a_2 Sun, Observables connected with the CP properties of the Higgs boson and defined with the momenta of the fermions}
\begin{align}
\label{the amplitude of X->Z_1* Z_2*}
A_{X \rightarrow Z_1^* Z_2^*} (\lambda_1, \lambda_2) = & 2 \sqrt {\sqrt{2} G_F} m_Z^2 \Bigl (a_Z  (e_1^* \cdot e_2^*) + \frac {b_Z} {m_X^2} (e_1^* \cdot (p_1 + p_2)) (e_2^* \cdot (p_1 + p_2)) + & \notag \\
& + i \frac {c_Z} {m_X^2} \varepsilon_{\mu \nu \rho \sigma} (p_1^{\mu} + p_2^{\mu}) (p_1^{\nu} - p_2^{\nu}) (e_1^{\rho})^* (e_2^{\sigma})^* \Bigr), &
\end{align}
where $\lambda_j$, $e_j$, $p_j$ are respectively the helicity, the polarization 4-vector and the 4-momentum of the boson $Z_j^*$, $G_F$ is the Fermi constant, $m_Z$ is the mass of the $Z$ boson, $a_Z$, $b_Z$, $c_Z$ are complex-valued dimensionless functions of $a_1$ and $a_2$, $\varepsilon_{\mu \nu \rho \sigma}$ is the Levi-Civita symbol ($\varepsilon_{0123} = 1$). Note that at tree level
\begin{itemize}
\item[---] if $X$ is the SM Higgs boson, then $a_Z = 1$, $b_Z = c_Z = 0$;
\item[---] if the $CP$ parity of $X$ is -1, then $a_Z = b_Z = 0$ and $c_Z \neq 0$;
\item[---] if the $CP$ parity of $X$ is indefinite, then $a_Z \neq 0$, $c_Z \neq 0$ and/or $b_Z \neq 0$, $c_Z \neq 0$.
\end{itemize}

Calculating the Lorentz-invariant amplitude $A_{X \rightarrow Z_1^* Z_2^*} (\lambda_1, \lambda_2)$ in a reference frame in which $\mathbf{p_1} + \mathbf{p_2} = 0$, we derive that
\begin{align}
\label{amplitudes of X->Z_1* Z_2* for various helicities of Z_1* and Z_2*}
& A_{X \rightarrow Z_1^* Z_2^*} (-1, -1) = 2 \sqrt {\sqrt{2} G_F} m_Z^2 \left (a_Z - c_Z \frac{\lambda^{\frac{1}{2}} (m_X^2, a_1, a_2)}{m_X^2} \right), & \notag \\
& A_{X \rightarrow Z_1^* Z_2^*} (0, 0) = -2 \sqrt {\sqrt{2} G_F} m_Z^2 \left (a_Z \frac{m_X^2 - a_1 - a_2}{2 \sqrt{a_1 a_2}} + b_Z \frac{\lambda (m_X^2, a_1, a_2)}{4 m_X^2 \sqrt{a_1 a_2}} \right), & \notag \\
& A_{X \rightarrow Z_1^* Z_2^*} (1, 1) = 2 \sqrt {\sqrt{2} G_F} m_Z^2 \left (a_Z + c_Z \frac{\lambda^{\frac{1}{2}} (m_X^2, a_1, a_2)}{m_X^2} \right), & \notag \\
& A_{X \rightarrow Z_1^* Z_2^*} (\lambda_1, \lambda_2) = 0, ~~ {\rm    если} ~~ \lambda_1 \neq \lambda_2, &
\end{align}
where the function $\lambda (x, y, z)$ is defined in the standard way: $\lambda (x, y, z) = x^2 + y^2 + z^2 - 2xy - 2xz - 2 yz$.

\begin{figure}[h]
\includegraphics[scale=0.7]{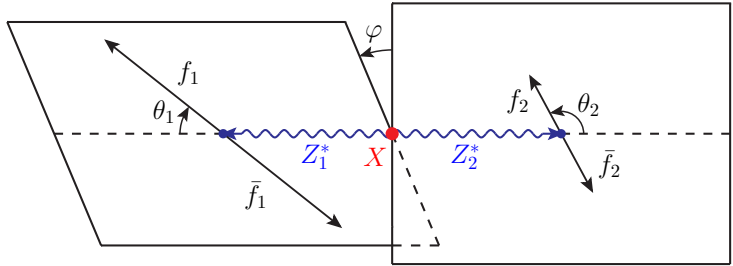}
\caption{The kinematics of the decay $X \rightarrow Z_1^* Z_2^* \rightarrow f_1 \bar{f}_1 f_2 \bar{f}_2$. The momenta of $Z_1^*$ and $Z_2^*$ are shown in a rest frame of $X$, the momenta of $f_1$ and $\bar{f}_1$ ($f_2$ and $\bar{f}_2$) are shown in a rest frame of $Z_1^*$ ($Z_2^*$).}
\label{momenta of particles produced in an XZZ decay}
\end{figure}

To describe the decay (\ref{X-> Z_1^* Z_2^* -> f_1 antif_1 f_2 antif_2}), let us introduce the following angles (see Fig.~\ref{momenta of particles produced in an XZZ decay}): $\theta_1$ ($\theta_2$) is the angle between the momentum of $Z_1^*$ ($Z_2^*$) in a rest frame of $X$ and the momentum of $f_1$ ($f_2$)  in a rest frame of $Z_1^*$ ($Z_2^*$) and $\varphi$ is the azimuthal angle between the planes of the decays $Z_1^* \rightarrow f_1 \bar{f}_1$ and $Z_2^* \rightarrow f_2 \bar{f}_2$. Further we go into the case of the non-identical fermions, $f_1 \neq f_2$. Using the helicity formalism (see, for example, \cite{the helicity formalism}), we obtain that in the approximation of the massless fermions, $m_{f_1} = m_{f_2} = 0$, the differential decay width of (\ref{X-> Z_1^* Z_2^* -> f_1 antif_1 f_2 antif_2}) with respect to $a_1$, $a_2$, $\theta_1$, $\theta_2$, $\varphi$ appears as follows:
\begin{align}
\label{the differential width with respect to a_1, a_2, theta_1, theta_2, varphi}
\frac{d^5 \Gamma} {da_1 da_2 d \theta_1 d \theta_2 d \varphi} = & \frac {\sqrt{2} G_F^3 m_Z^8} {(4 \pi)^6 m_X^3} (a_{f_1}^2 + v_{f_1}^2) (a_{f_2}^2 + v_{f_2}^2) \frac{\lambda^{\frac{1}{2}} (m_X^2, a_1, a_2) a_1 a_2}
{ D(a_1) D(a_2)}  & \notag \\
& \times \sin \theta_1 \sin \theta_2 [(|A_{\parallel}|^2 + |A_{\perp}|^2) \left ((1 + \cos^2 \theta_1) (1 + \cos^2 \theta_2)
 + 4 A_{f_1} A_{f_2} \cos \theta_1 \cos \theta_2 \right)  & \notag \\
& + 4 |A_0|^2 \sin^2 \theta_1 \sin^2 \theta_2 - 4 \, {\rm Re}(A_{\parallel}^* A_{\perp}) (A_{f_1} \cos \theta_1 (1 + \cos^2 \theta_2) + A_{f_2} \cos \theta_2 (1 + \cos^2 \theta_1))  & \notag \\
& + 4 \sqrt{2} \sin \theta_1 \sin \theta_2 (({\rm  Re}(A_0^* A_{\parallel}) \cos \varphi - {\rm Im}(A_0^* A_{\perp}) \sin \varphi) (A_{f_1} A_{f_2} + \cos \theta_1 \cos \theta_2)  & \notag \\
& - ({\rm  Re}(A_0^* A_{\perp}) \cos \varphi - {\rm Im}(A_0^* A_{\parallel}) \sin \varphi) (A_{f_1} \cos \theta_2 + A_{f_2} \cos \theta_1))  & \notag \\
& + \sin^2 \theta_1 \sin^2 \theta_2 ((|A_{\parallel}|^2 - |A_{\perp}|^2) \cos 2 \varphi - 2 \, {\rm Im}(A_{\parallel}^* A_{\perp}) \sin 2 \varphi) ], &
\end{align}
where $a_f$ is the projection of the weak isospin of a fermion $f$, $v_f \equiv a_f - 2 \frac{q_f}{e} \sin^2 \theta_W$, $q_f$ is the electric charge of the fermion $f$, $e$ is the electric charge of the positron, $\theta_W$ is the weak mixing angle, $D(a_{1,2}) \equiv (a_{1,2}- m_Z^2)^2 + (m_Z \Gamma_Z)^2$, $\Gamma_Z$ is the total width of the $Z$ boson, $A_f \equiv \frac{2 a_f v_f}{a_f^2 + v_f^2}$,
\begin{align}
\label{definitions and formulae for A_0, A_parallel, A_perp for a decay X->Z_1^* Z_2^*}
& A_{\parallel} (a_1, a_2) \equiv \frac{A_{X \rightarrow Z_1^* Z_2^*} (1, 1) + A_{X \rightarrow Z_1^* Z_2^*} (-1, -1)}{2^{\frac{7}{4}} \sqrt {G_F} m_Z^2} = \sqrt{2} a_Z, & \notag \\
& A_{\perp} (a_1, a_2) \equiv \frac{A_{X \rightarrow Z_1^* Z_2^*} (1, 1) - A_{X \rightarrow Z_1^* Z_2^*} (-1, -1)}{2^{\frac{7}{4}} \sqrt {G_F} m_Z^2} = \sqrt{2} c_Z \frac{\lambda^{\frac{1}{2}} (m_X^2, a_1, a_2)}{m_X^2}, & \notag \\
& A_0 (a_1, a_2) \equiv \frac{A_{X \rightarrow Z_1^* Z_2^*} (0, 0)}{2^{\frac{5}{4}} \sqrt {G_F} m_Z^2} = - \left (a_Z \frac{m_X^2 - a_1 - a_2}{2 \sqrt{a_1 a_2}} + b_Z \frac{\lambda (m_X^2, a_1, a_2)}{4 m_X^2 \sqrt{a_1 a_2}} \right). &
\end{align}

Futher the approximation $m_{f_1} = m_{f_2} = 0$ is used. Using Eq.~(\ref{the differential width with respect to a_1, a_2, theta_1, theta_2, varphi}), one can connect the ratios of quantities $|A_0|^2$, $|A_{\parallel}|^2 + |A_{\perp}|^2$, $|A_{\parallel}|^2 - |A_{\perp}|^2$,  ${\rm   Re} (A_0^* A_{\parallel})$, ${\rm Im}(A_0^* A_{\parallel})$, ${\rm  Re}(A_0^* A_{\perp})$, ${\rm Im}(A_0^* A_{\perp})$, ${\rm  Re}(A_{\parallel}^* A_{\perp})$, ${\rm Im}(A_{\parallel}^* A_{\perp})$ to $|A_0|^2 + |A_{\parallel}|^2 + |A_{\perp}|^2$ with functions of $a_1$, $a_2$ which can be measured in experiment. We will call these ratios the helicity coefficients of the decay $X \rightarrow Z_1^* Z_2^*$.


\subsection{A differential width $\frac {d^2 \Gamma} {da_1 da_2}$}
\label{subsec:diff_width with respect to a_1, a_2}

The number of the decays
\begin{align}
\label{h->Z_1^* Z_2^*->4l}
h \rightarrow Z_1^* Z_2^* \rightarrow l_1^-  l_1^+  l_2^-  l_2^+ ~~ (l_j = e, \mu),
\end{align}
detected in the ATLAS experiment \cite{h->Z_1^* Z_2^*->4l and h->W-* W+*->l nu l nu ATLAS} wherein the invariant mass of the four leptons was in the interval [120 GeV, 130 GeV], is equal to 32. The number of the decays (\ref{h->Z_1^* Z_2^*->4l}) detected in the CMS experiment \cite{h->Z_1^* Z_2^*->4l CMS} in which the four-lepton invariant mass was within [121.5 GeV, 130.5 GeV], is equal to 25. In view of the insignificant amount of data, at the present time an experimental dependence of the distribution $\frac{1}{\Gamma} \frac{d^5 \Gamma} {da_1 da_2 d \theta_1 d \theta_2 d \varphi}$ ($\Gamma$ is the total width of the decay (\ref{X-> Z_1^* Z_2^* -> f_1 antif_1 f_2 antif_2})) for any of the decays (\ref{h->Z_1^* Z_2^*->4l}) is not available. Let us consider differential decay widths of (\ref{X-> Z_1^* Z_2^* -> f_1 antif_1 f_2 antif_2}) with respect to four and fewer variables. Integrating Eq. (\ref{the differential width with respect to a_1, a_2, theta_1, theta_2, varphi}) with respect to $\theta_1$, $\theta_2$, $\varphi$, we obtain
\begin{align}
\label{the differential width with respect to a_1, a_2}
\frac {d^2 \Gamma} {da_1 da_2} = \frac {\sqrt{2} G_F^3 m_Z^8} {9 (2 \pi)^5 m_X^3} (a_{f_1}^2 + v_{f_1}^2) (a_{f_2}^2 + v_{f_2}^2)  \frac{\lambda^{\frac{1}{2}} (m_X^2, a_1, a_2) a_1 a_2} { D(a_1) D(a_2)} \sum_{p = 0,\parallel, \perp} |A_p|^2.
\end{align}
It follows from Eqs. (\ref{the differential width with respect to a_1, a_2}), (\ref{definitions and formulae for A_0, A_parallel, A_perp for a decay X->Z_1^* Z_2^*}) that the dependence of the differential width $\frac {d^2 \Gamma} {da_1 da_2}$ on $a_Z$, $b_Z$, $c_Z$ boils down only to the dependence on $|a_Z|$, $|b_Z|$, $|c_Z|$ and on $\cos{(\arg b_Z - \arg a_Z)}$.

The available experimental data on properties of the particle $h$ are close to the corresponding theoretical predictions about the SM Higgs boson (see, for example, \cite{h->Z_1^* Z_2^*->4l and h->W-* W+*->l nu l nu ATLAS, h->Z_1^* Z_2^*->4l CMS, the probability that CPh=-1 is very small CMS}). That is why $a_{hZ} \approx 1$, $b_{hZ} \approx 0$, $c_{hZ} \approx 0$, where
$$ a_{hZ} \equiv a_Z |_{X = h}, ~b_{hZ} \equiv b_Z |_{X = h}, ~c_{hZ} \equiv c_Z |_{X = h}.$$

In Fig.~\ref{a plot of the differential width with respect to a_1, a_2 in the SM} we show the differential decay width (\ref{the differential width with respect to a_1, a_2}) for $X \rightarrow Z_1^* Z_2^* \rightarrow l_1^-  l_1^+  l_2^-  l_2^+ ~~ (l_j = e, \mu, \tau, l_1 \neq l_2)$ as a function of $\sqrt{a_1}$, $\sqrt{a_2}$ in the SM for $|a_Z| = 1$, $b_Z = c_Z = 0$ and $m_X = m_h$, where $m_h$ is the mass of the Higgs boson $h$.
The range of $\sqrt{a_1}$, $\sqrt{a_2}$ in this plot is determined by the inequalities (\ref{intervals of a_1 and a_2}) in the approximation of the massless fermions. In calculations and when plotting graphs the experimental data listed in Table~\ref{experimental values of 'constants' related to decays X-> Z_1^* Z_2^* -> f_1 antif_1 f_2 antif_2} are used, and $\sin^2 \theta_W = 1 - m_W^2 / m_Z^2$, where $m_W$ is the mass of the $W$ boson.

\begin{figure}[h]
\includegraphics[scale=0.7]{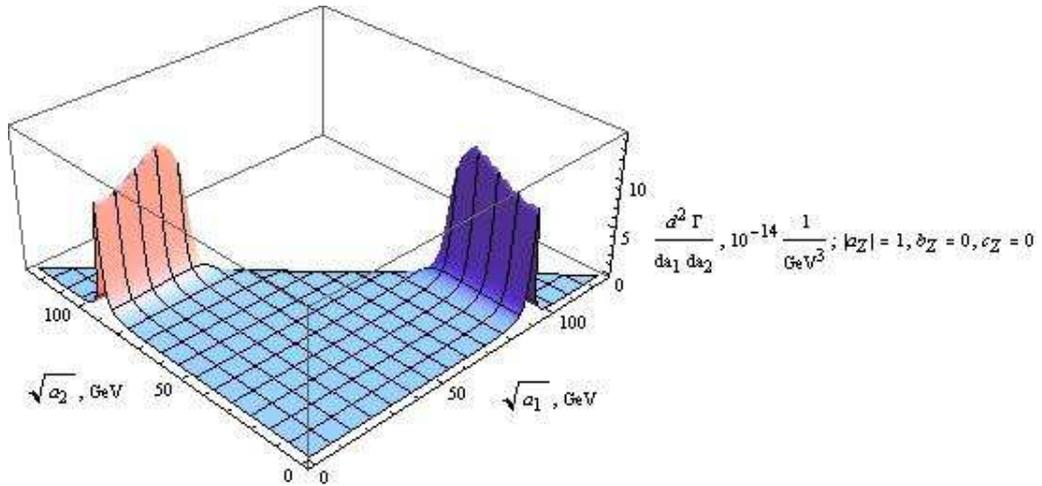}
\caption{The dependence of the differential decay width $\frac {d^2 \Gamma} {da_1 da_2}$ of $X \rightarrow Z_1^* Z_2^* \rightarrow l_1^- l_1^+  l_2^-  l_2^+ ~~ (l_j = e, \mu, \tau, l_1 \neq l_2)$ on $\sqrt{a_1}$ and $\sqrt{a_2}$ in the SM for $m_X = m_h$.}
\label{a plot of the differential width with respect to a_1, a_2 in the SM}
\end{figure}

\begin{table}[h]
\caption{Values of the Fermi constant, of the masses of $h$, $Z$, $W$ and of the total width of $Z$ \cite{Particle Data Group 2014}.}
\label{experimental values of 'constants' related to decays X-> Z_1^* Z_2^* -> f_1 antif_1 f_2 antif_2}
\begin{tabular}{l}
\hline \hline
$G_F = 1.1663787(6) \times 10^{-5} ~{\rm GeV}^{-2}$ \\
$m_h = 125.7(4) ~{\rm GeV}$ \\
$m_Z = 91.1876(21) ~{\rm GeV}$ \\
$m_W = 80.385(15) ~{\rm GeV}$ \\
$\Gamma_Z = 2.4952(23) ~{\rm GeV}$ \\
\hline \hline
\end{tabular}
\end{table}

As one can see from Fig.~\ref{a plot of the differential width with respect to a_1, a_2 in the SM}, in the SM the function $\frac {d^2 \Gamma} {da_1 da_2}$ has peaks at $\sqrt{a_1} = m_Z$ and $\sqrt{a_2} = m_Z$, resulting from the quantities $D(a_1)$ and $D(a_2)$ in (\ref{the differential width with respect to a_1, a_2}).

Let us calculate the ratio of a typical value of $\frac {d^2 \Gamma} {da_1 da_2}$ in the SM on the peaks to its typical value in an area in which $\sqrt{a_1}$ and $\sqrt{a_2}$ significantly differ from $m_Z$ (we will call this area ``plateau''). As indicative values of $\sqrt{a_1}$ and $\sqrt{a_2}$ on the peaks we take $\sqrt{a_1} = m_Z, \sqrt{a_2} = \frac{1}{2} (m_h - m_Z)$ and $\sqrt{a_1} = \frac{1}{2} (m_h - m_Z), \sqrt{a_2} = m_Z$ (see (\ref{intervals of a_1 and a_2})), and values on the ``plateau'' are chosen $\sqrt{a_1} = \sqrt{a_2} = \frac{1}{2} m_Z$.
It follows from (\ref{the differential width with respect to a_1, a_2}) that in the SM for any $f_1, \, f_2$
values of $\frac {d^2 \Gamma} {da_1 da_2}$ at $\sqrt{a_1} = m_Z$ or $\sqrt{a_2} = m_Z$ are approximately 100 times as great as values of this function on the ``plateau''.

If $m_X \neq m_h$ but just greater than $m_Z $, then $\sqrt{a_1}$ and/or $\sqrt{a_2}$ can be equal to $m_Z$ (according to (\ref{intervals of a_1 and a_2})), and, consequently, in this case the behavior of the function $\frac {d^2 \Gamma} {da_1 da_2}$ in the SM is similar to that in case $m_X = m_h$. That is why for any $m_X > m_Z$ and for any final fermions the differential width $\frac {d^2 \Gamma} {da_1 da_2}$ in the SM has a sharp maximum at $\sqrt{a_1} = m_Z$ or $\sqrt{a_2} = m_Z$. Therefore, if $|a_Z| \approx 1$, $b_Z \approx 0$, $c_Z \approx 0$ (which is the case of a small distinction between the couplings and their SM values), $\frac {d^2 \Gamma} {da_1 da_2}$ also has a sharp maximum at $\sqrt{a_1} = m_Z$ or $\sqrt{a_2} = m_Z$, provided that $m_X > m_Z$.


\subsection{Limits of applicability of the narrow-$Z$-width approximation}
\label{subsection: limits of applicability of the narrow-Z-width approximation}

In Refs. \cite{the narrow-width approximation is non-applicable for calculation of certain total widths, the narrow-width approximation is non-applicable for calculation of certain cross sections, a study of the narrow-width approximation} the accuracy of the narrow-width approximation has been studied for calculation of the total widths of various decays along with the total and differential cross sections of various processes. It is shown that in many cases (especially for processes beyond the SM) this approximation is not applicable. In this connection the question arises whether the narrow-$Z$-width approximation is applicable for obtaining the differential width $\frac{d \Gamma}{da_2}$ by means of integrating $\frac {d^2 \Gamma} {da_1 da_2}$. In this subsection we find the interval of all the $a_2$-values for which the approximate integration is valid.

We consider the $m_X$-values such that $m_X > m_Z $ and the dependences of $a_Z (a_1, a_2)$, $b_Z (a_1, a_2)$, $c_Z (a_1, a_2)$ such that for any $f_1$ and $f_2$ $\frac {d^2 \Gamma} {da_1 da_2}$ has a sharp maximum when $\sqrt{a_1} = m_Z$ or $\sqrt{a_2} = m_Z$ (an example of such dependences is $|a_Z| \approx 1$, $b_Z \approx 0$, $c_Z \approx 0$). Then while calculating the differential width $\frac {d \Gamma} {da_2}$ one may use the narrow-$Z$-width approximation:
\begin{eqnarray}
\label{an approximation of the delta function for the differential width with respect to a_2}
\frac {d \Gamma} {da_2} &=& \int \limits_{0}^{(m_X - \sqrt{a_2})^2} da_1 \frac {d^2 \Gamma} {da_1 da_2} \approx \int \limits_{0}^{(m_X - \sqrt{a_2})^2} da_1 \frac{\pi}{m_Z \Gamma_Z} \delta (a_1 - m_Z^2) f (a_1, a_2) \notag \\
&=& \frac{\pi}{m_Z \Gamma_Z} f (m_Z^2, a_2) \qquad \forall ~\sqrt{a_2} \in \left (0, \, m_X - m_Z - \Delta \right],
\end{eqnarray}
where $\Delta$ is some positive quantity and
\begin{align}
\label{a definition of f}
f (a_1, a_2) \equiv \frac {\sqrt{2} G_F^3 m_Z^8} {9 (2 \pi)^5 m_X^3} (a_{f_1}^2 + v_{f_1}^2) (a_{f_2}^2 + v_{f_2}^2)  \frac{\lambda^{\frac{1}{2}} (m_X^2, a_1, a_2) a_1 a_2} {D (a_2)}
\sum_{p = 0,\parallel, \perp} |A_p|^2.
\end{align}
$\Delta > 0$ since in Eq.~(\ref{an approximation of the delta function for the differential width with respect to a_2}) one may use the approximation $\frac {d^2 \Gamma} {da_1 da_2} = \frac{\pi}{m_Z \Gamma_Z} \delta (a_1 - m_Z^2) f (a_1, a_2)$
only when $\sqrt{a_2} < m_X - m_Z$, because if $\sqrt{a_2}$ approaches $m_X - m_Z$, the peak of $\frac {d^2 \Gamma} {da_1 da_2}$ at $\sqrt{a_1} = m_Z$ gets less sharp and at $\sqrt{a_2} = m_X - m_Z$ the peak disappears (see Fig.~\ref{a plot of the differential width with respect to a_1, a_2 in the SM} and Eq.~(\ref{the differential width with respect to a_1, a_2})). However, the derivation (\ref{an approximation of the delta function for the differential width with respect to a_2}) does not allow one to estimate the accuracy of the formula $\frac {d \Gamma} {da_2} \approx \frac{\pi}{m_Z \Gamma_Z} f (m_Z^2, a_2)$ at a given value of $\sqrt{a_2}$, and for this reason it is not clear what value of $\Delta$ should be chosen.

To clarify this point, let us derive the formula for $\frac {d \Gamma} {da_2}$ in the following way:

\begin{align}
\label{approximate formulas for the differential width with respect to a_2}
 \frac {d \Gamma} {da_2} & = \int \limits_{0}^{(m_X - \sqrt{a_2})^2} da_1 \frac {d^2 \Gamma} {da_1 da_2} \approx \int \limits_{m_Z^2 - \varepsilon_1}^{m_Z^2 + \varepsilon_2} da_1 \frac {d^2 \Gamma} {da_1 da_2} = \int \limits_{m_Z^2 - \varepsilon_1}^{m_Z^2 + \varepsilon_2} da_1 \frac{f (a_1, a_2)}{(a_1- m_Z^2)^2 + (m_Z \Gamma_Z)^2} & \notag \\
& \approx \int \limits_{m_Z^2 - \varepsilon_1}^{m_Z^2 + \varepsilon_2} da_1 \frac{f (m_Z^2, a_2)}{(a_1- m_Z^2)^2 + (m_Z \Gamma_Z)^2} = \frac{\arctan \frac{\varepsilon_2}{m_Z \Gamma_Z} + \arctan \frac{\varepsilon_1}{m_Z \Gamma_Z}} {m_Z \Gamma_Z} f (m_Z^2, a_2) & \notag \\
& \approx \frac{\pi}{m_Z \Gamma_Z} f (m_Z^2, a_2), &
\end{align}
where $\varepsilon_1$ and $\varepsilon_2$ are some positive quantities such that $m_Z \Gamma_Z \ll \varepsilon_j \ll m_Z^2$, the variable $a_2$ takes values in the interval $\left (0, \left (m_X - \sqrt{m_Z^2 + \varepsilon_2} \right)^2 \right]$.

One of the approximations used in Eq.~(\ref{approximate formulas for the differential width with respect to a_2}) is the switch from the integration over an interval $(0, (m_X - \sqrt{a_2})^2)$ to the integration over an interval $(m_Z^2 - \varepsilon_1, m_Z^2 + \varepsilon_2)$. Thus, $m_Z^2 - \varepsilon_1$ has to be greater than or equal to $4 m_{f_1}^2$ (which holds true since $\varepsilon_1 \ll m_Z^2$) and $m_Z^2 + \varepsilon_2$ has to be  less than or equal to $(m_X - \sqrt{a_2})^2$, i.e. $a_2 \leq \left (m_X - \sqrt{m_Z^2 + \varepsilon_2} \right)^2$. The latter inequality restricts the interval of all the $a_2$-values for which these approximations are applicable. Consequently, in order to apply them for as long an interval of $a_2$-values as possible, one should use the minimal $\varepsilon_2$-value at which the approximations are valid.

While obtaining (\ref{approximate formulas for the differential width with respect to a_2}), we also used an approximation $A \approx \pi$ \ ($A \equiv \arctan \frac{\varepsilon_2}{m_Z \Gamma_Z} + \arctan \frac{\varepsilon_1}{m_Z \Gamma_Z}$). Let us define $\varepsilon_1$ as $\varepsilon_1 \equiv m_Z \sqrt{m_Z \Gamma_Z}$ (so as $\frac{\varepsilon_1}{m_Z \Gamma_Z} = \frac{m_Z^2}{\varepsilon_1}$). The values of quantities $A$ and $m_h - \sqrt{m_Z^2 + \varepsilon_2}$ which are listed in Table~\ref{applicability of the narrow-Z-width approximation} specify for the considered  $\varepsilon_2$-values the accuracy of the approximation $A \approx \pi$ and the maximal value of $\sqrt{a_2}$ at which the narrow-$Z$-width approximation is applicable in case $X = h$.

\begin{table}[h]
\caption{Values of $A$ ($\varepsilon_1 \equiv m_Z \sqrt{m_Z \Gamma_Z}$) and of $m_h - \sqrt{m_Z^2 + \varepsilon_2}$ at various values of $\varepsilon_2$.}
\label{applicability of the narrow-Z-width approximation}
\begin{tabular}{c c c}
\hline \hline
$\varepsilon_2$ & $A$ & $m_h - \sqrt{m_Z^2 + \varepsilon_2}$ ({\rm GeV}) \\
\hline
0 & $0.45 \pi$ & 34.51 \\
$m_Z \Gamma_Z$ & $0.70 \pi$ & 33.27 \\
$2 m_Z \Gamma_Z$ & $0.80 \pi$ & 32.05 \\
$3 m_Z \Gamma_Z$ & $0.85 \pi$ & 30.84 \\
$4 m_Z \Gamma_Z$ & $0.87 \pi$ & 29.65 \\
\hline \hline
\end{tabular}
\end{table}

According to Table~\ref{applicability of the narrow-Z-width approximation}, if $\varepsilon_2 < 3 m_Z \Gamma_Z$, then $A < 0.85 \pi$ and, in view of the big difference between $A$ and $\pi$, we will not apply the approximations (\ref{approximate formulas for the differential width with respect to a_2}) for such values of $\varepsilon_2$. Hence we will use $\varepsilon_2 = 3 m_Z \Gamma_Z$. It follows from (\ref{approximate formulas for the differential width with respect to a_2}) that
\begin{align}
\label{the final approximate formula for the differential width with respect to a_2}
& \frac {d \Gamma} {da_2} \approx \frac {\sqrt{2} G_F^3 m_Z^9} {9 \cdot 2^5 \pi^4 m_X^3 \Gamma_Z} (a_{f_1}^2 + v_{f_1}^2) (a_{f_2}^2 + v_{f_2}^2) \frac{\lambda^{\frac{1}{2}} (m_X^2, m_Z^2, a_2) a_2} {D (a_2)}
\sum_{p = 0,\parallel, \perp} |A_p^\prime|^2 &  \\
& \forall ~\sqrt{a_2} \in \left(0, \, m_X - \sqrt{m_Z^2 + \varepsilon_2} \right], & \notag
\end{align}
where $A_p^{\prime} \equiv A_p (m_Z^2, a_2)$ ($p = 0, \parallel, \perp$).

Note that in Refs.~\cite{distributions of a_2 Choi, distributions of a_2 Menon, distributions of a_2 Sun} when plotting dependences of $\frac{1}{\Gamma} \frac {d \Gamma} {da_2}$ on $\sqrt{a_2}$, formulas for $\frac {d \Gamma} {da_2}$ which correspond to (\ref{the final approximate formula for the differential width with respect to a_2}) have been used, but these graphs have been plotted for $\sqrt{a_2} \leq m_X - m_Z$, despite the fact that Eq.~(\ref{approximate formulas for the differential width with respect to a_2}) is not valid at $\varepsilon_2 = 0$ (see Table~\ref{applicability of the narrow-Z-width approximation}), and, therefore, the plotted dependences significantly differ from the true ones in the interval $\sqrt{a_2} \in (m_X - \sqrt{m_Z^2 + 3 m_Z \Gamma_Z}, m_X - m_Z]$.


\subsection{An inequality constraining $a_{hZ}^{\prime}$, $b_{hZ}^{\prime}$, $c_{hZ}^{\prime}$ from CMS data}
\label{subsection: an inequality constraining a, b, c from CMS data}

According to \cite{h->Z_1^* Z_2^*->4l CMS},
\begin{align}
\label{(sigma by the branching) over (sigma in the SM by the branching in the SM)}
\frac{\sigma (pp \rightarrow h) \frac{\Gamma (h \rightarrow Z_1^* Z_2^* \rightarrow 4l)}{\Gamma_h}} {\sigma_{SM} (pp \rightarrow h) \frac{\Gamma_{SM} (h \rightarrow Z_1^* Z_2^* \rightarrow 4l)}{\Gamma_{h \, SM}}} = 0.93_{-0.23}^{+0.26}({\rm stat})_{-0.09}^{+0.13}({\rm syst}),
\end{align}
where $\sigma (pp \rightarrow h)$ is the cross section for production of $h$ in $pp$ collisions,
\begin{align}
\Gamma (h \rightarrow Z_1^* Z_2^* \rightarrow 4l) & \equiv \Gamma (h \rightarrow Z_1^* Z_2^* \rightarrow 4e) + \Gamma (h \rightarrow Z_1^* Z_2^* \rightarrow 4 \mu) + \Gamma (h \rightarrow Z_1^* Z_2^* \rightarrow 2e 2\mu) & \notag \\
& = 2 \, \Gamma (h \rightarrow Z_1^* Z_2^* \rightarrow 4e) + \Gamma (h \rightarrow Z_1^* Z_2^* \rightarrow 2e 2\mu), &
\end{align}
$\Gamma_h$ is the total width of the boson $h$, $\sigma_{SM} (pp \rightarrow h)$, $\Gamma_{SM} (h \rightarrow Z_1^* Z_2^* \rightarrow 4 l)$, $\Gamma_{h \, SM}$ are the predictions of the SM for respectively $\sigma (pp \rightarrow h)$, $\Gamma (h \rightarrow Z_1^* Z_2^* \rightarrow 4 l)$, $\Gamma_h$ at $m_h$ = 125.6 GeV. Obtaining (\ref{(sigma by the branching) over (sigma in the SM by the branching in the SM)}), the CMS collaboration has combined data from $pp$ collisions corresponding to an integrated luminosity of 5.1 fb$^{-1}$ at a center-of-mass energy $\sqrt{s}$ = 7 TeV and 19.7 fb$^{-1}$ at $\sqrt{s}$ = 8 TeV.

We consider the case in which the functions $|a_{hZ}^{\prime}|$, $|b_{hZ}^{\prime}|$, $|c_{hZ}^{\prime}|$, $\cos (\arg b_{hZ}^{\prime} - \arg a_{hZ}^{\prime})$ do not depend on $a_2$. Here we define
\[a_{hZ}^{\prime} \equiv a_{hZ} (m_Z^2, a_2), ~b_{hZ}^{\prime} \equiv b_{hZ} (m_Z^2, a_2), ~c_{hZ}^{\prime} \equiv c_{hZ} (m_Z^2, a_2). \]
Then using the approximation
\begin{align}
\label{sigma (pp->h) by Gamma_h approx sigma_SM (pp->h) by Gamma_h SM}
\frac{\sigma (pp \rightarrow h)} {\Gamma_h} \approx \frac{\sigma_{SM} (pp \rightarrow h)} {\Gamma_{h \, SM}}
\end{align}
and Eqs.~(\ref{(sigma by the branching) over (sigma in the SM by the branching in the SM)}) (within one standard deviation), (\ref{a formula for Gamma when a_Z, b_Z, c_Z are constants and f_1 and f_2 are arbitrary}), (\ref{the explicit dependence of f_0 on a, b, c in case m_X = m_h}) (see Appendix \ref{calculation of the total widths of decays X->Z_1^* Z_2^*->f_1 antif_1 f_2 antif_2 and h->Z_1^* Z_2^*}), we derive the relation
\begin{align}
\label{a constraint on a, b, c according to experimental data of the CMS}
|a_{hZ}^{\prime}|^2 + 0.015 \, |b_{hZ}^{\prime}|^2 + 0.177 \, {\rm Re} (a_{hZ}^{\prime *} b_{hZ}^{\prime}) + 0.037 \, |c_{hZ}^{\prime}|^2\, \in \, [0.68, \, 1.22].
\end{align}
While obtaining (\ref{a constraint on a, b, c according to experimental data of the CMS}) we plugged the central values of $m_h$, $m_Z$, $\Gamma_Z$ listed in Table~\ref{experimental values of 'constants' related to decays X-> Z_1^* Z_2^* -> f_1 antif_1 f_2 antif_2} into Eq.~(\ref{a formula for Gamma when a_Z, b_Z, c_Z are constants and f_1 and f_2 are arbitrary}). Note that the latter equation is derived at tree level and without allowance for the interference term connected with the permutation of the identical fermions in case $f_1 = f_2$. The interference contribution to $\Gamma (h \rightarrow Z_1^* Z_2^* \rightarrow 4l)$ at tree level is expected to be negligible since in the SM at $m_h = 140$ GeV it amounts to 2.99\% (see Table 1 in Ref.~\cite{The contribution of the interference term to Gamma (h to ZZ to 4e) in the SM at tree level}). Using the data of Table~\ref{sigma (pp->h) and Gamma_h in experiment and theory} and considering two sigma errors where available, we obtain that at $\sqrt{s}$ = 8 TeV
\begin{align}
\label{a `two sigma' interval of sigma (pp->h) by Gamma_h}
\frac{\sigma (pp \rightarrow h) / \Gamma_h} {\sigma_{SM} (pp \rightarrow h) / \Gamma_{h \, SM}} \in (0.17, \infty),
\end{align}
which means that the approximation (\ref{sigma (pp->h) by Gamma_h approx sigma_SM (pp->h) by Gamma_h SM}) does not contradict the experimental limits.

\begin{table}[h]
\caption{Experimental and theoretical results for the total production cross-section of the Higgs boson in $pp$ collisions and for its total width.}
\label{sigma (pp->h) and Gamma_h in experiment and theory}
\begin{tabular}{l}
\hline \hline
$\sigma (pp \rightarrow h) = 33.0 \pm 5.3 ({\rm stat}) \pm 1.6 ({\rm syst})$ pb at $\sqrt{s}$ = 8 TeV \cite{The total and differential cross sections of the process pp to h ATLAS} \\
$\sigma_{SM} (pp \rightarrow h) = 22.09$ pb (uncertainties not available) at $\sqrt{s}$ = 8 TeV \cite{LHC 2012-2014 data on SM Higgs boson branchings and total cross sections} \\
$\Gamma_h$ < 22 MeV at 95\% confidence level (CL) \cite{Constraints on the total width of the Higgs boson CMS} \\
$\Gamma_{h \, SM} = 4.15 \pm 0.16$ MeV \cite{Properties of the Higgs boson (theory and experiment)} \\
\hline \hline
\end{tabular}
\end{table}

Moreover, assuming that all the couplings of the Higgs boson except for $a_{hZ}$, $b_{hZ}$ and $c_{hZ}$ are equal to their SM values, we can verify (\ref{sigma (pp->h) by Gamma_h approx sigma_SM (pp->h) by Gamma_h SM}). In this case the only anomalous contribution to $\Gamma_h$ comes from $\Gamma(h \to Z_1^* Z_2^*)$, which makes up, in the SM, only about 2.81\% \cite{Properties of the Higgs boson (theory and experiment)} of the total Higgs boson width, and therefore $\Gamma_h$ is unlikely to substantially differ from its SM prediction. Besides, the inequality (\ref{a constraint on a, b, c according to experimental data of the CMS}) means that $|\Gamma (h \rightarrow Z_1^* Z_2^*) - \Gamma_{SM} (h \rightarrow Z_1^* Z_2^*)| / \Gamma_{SM} (h \rightarrow Z_1^* Z_2^*) \in [0, 0.32]$ because its left-hand side is
\[ \frac{\Gamma (h \rightarrow Z_1^* Z_2^* \rightarrow 4 l)} {\Gamma_{SM} (h \rightarrow Z_1^* Z_2^* \rightarrow 4 l)} = \frac{\Gamma (h \rightarrow Z_1^* Z_2^*)} {\Gamma_{SM} (h \rightarrow Z_1^* Z_2^*)}\]
(see (\ref{(sigma by the branching) over (sigma in the SM by the branching in the SM)}), (\ref{a formula for Gamma when a_Z, b_Z, c_Z are constants and f_1 and f_2 are arbitrary})). For this reason (\ref{a constraint on a, b, c according to experimental data of the CMS}) implies that the relative change of $\Gamma_h$ is less than $2.81\% \cdot 0.32 \approx 0.90\%$, and, consequently, (\ref{a constraint on a, b, c according to experimental data of the CMS}) is consistent with the approximation $\Gamma_h \approx \Gamma_{h \, SM}$.

The dominant contribution to the Higgs boson production cross section $\sigma_{SM} (pp \rightarrow h)$ comes from the gluon fusion process $g g \to h$, which is independent of the $h ZZ$ vertex. The processes involving the $hZZ$ interaction, i.e. the Higgs-strahlung $Zh$ and the $Z$ boson fusion, constitute much less parts of $\sigma_{SM} (pp \rightarrow h)$. Specifically, at $\sqrt{s}$ = 8 TeV they can be estimated as 0.41 pb and 0.70 pb respectively \cite{LHC 2012-2014 data on SM Higgs boson branchings and total cross sections}. The total production cross section at this energy is 22.09 pb (see Table~\ref{sigma (pp->h) and Gamma_h in experiment and theory}), so the processes of interest contribute about 5\% of the total cross section. That is why it seems improbable that the couplings $a_{hZ}$, $b_{hZ}$ and $c_{hZ}$ provide a significant difference between $\sigma (pp \rightarrow h)$ and $\sigma_{SM} (pp \rightarrow h)$. However, a derivation of the dependence of the total production cross section on the $hZZ$ couplings would require a separate study.

Summarizing the discussion of the approximation (\ref{sigma (pp->h) by Gamma_h approx sigma_SM (pp->h) by Gamma_h SM}), we can infer, firstly, that it is consistent with the available data~\cite{The total and differential cross sections of the process pp to h ATLAS,Constraints on the total width of the Higgs boson CMS} and, secondly, under the assumption that the only anomalous Higgs boson couplings are related to the $h ZZ$ vertex, Eq.~(\ref{sigma (pp->h) by Gamma_h approx sigma_SM (pp->h) by Gamma_h SM}) is most likely to be valid due to the small contributions of the $hZZ$ vertex to $\Gamma_h$ and $\sigma (pp \rightarrow h)$.

\subsection{Constraints on $a_{hZ}^{\prime}$, $b_{hZ}^{\prime}$, $c_{hZ}^{\prime}$}
\label{subsection: constraints on a, b, c}

The inequality (\ref{a constraint on a, b, c according to experimental data of the CMS}) constrains the whole six-dimensional space formed by the real and imaginary parts of the couplings $a_{hZ}^{\prime}$, $b_{hZ}^{\prime}$ and $c_{hZ}^{\prime}$ to the set of ellipsoids allowed by (\ref{(sigma by the branching) over (sigma in the SM by the branching in the SM)}). Note that a similar interpretation has been suggested in Ref.~\cite{an approach to getting constraints on the XZZ couplings by means of finding the couplings as a point on a sphere}.

From (\ref{a constraint on a, b, c according to experimental data of the CMS}) it follows that the variant $a_{hZ}^{\prime} = 0, b_{hZ}^{\prime} = 0, c_{hZ}^{\prime} = 1$ (negative $CP$ parity of the boson $h$) is excluded. Now let us find constraints on the values of $b_{hZ}^{\prime}$ and $c_{hZ}^{\prime}$, assuming that $a_{hZ}^{\prime}$ is taken from the SM, i.e. $|a_{hZ}^{\prime}| = 1$ or $a_{hZ}^{\prime} = 1$. Then
\begin{subequations}
\label{possible sets of values of a, b, c}
\begin{align}
\label{an allowed interval of c at |a| = 1 and b = 0}
& |a_{hZ}^{\prime}| = 1 ~~ {\rm and} ~~ b_{hZ}^{\prime} = 0 \, \Rightarrow \, |c_{hZ}^{\prime}| \in [0, \, 2.44]; \\
\label{an allowed interval of b at a = 1 and c = 0 and Im b = 0}
& a_{hZ}^{\prime} = 1 ~~ {\rm and} ~~ c_{hZ}^{\prime} = 0 ~~ {\rm and} ~~ {\rm  Im}\,b_{hZ}^{\prime} = 0 \, \Rightarrow \, b_{hZ}^{\prime} \in ([-12.66, \, -9.31] \cup [-2.22, \, 1.14]), \\
\label{an allowed interval of Im b at a = 1 and c = 0 and Re b = 0}
& a_{hZ}^{\prime} = 1 ~~ {\rm and} ~~ c_{hZ}^{\prime} = 0 ~~ {\rm and} ~~ {\rm  Re} \,b_{hZ}^{\prime} = 0 \, \Rightarrow \, {\rm  Im}\,b_{hZ}^{\prime} \in [-3.84, \, 3.84].
\end{align}
\end{subequations}

Let us compare (\ref{possible sets of values of a, b, c}) with the $hZZ$ coupling constraints obtained by the CMS \cite {Probabilities of Higgs boson spin-parity hypotheses and constraints on the hVV couplings CMS} and ATLAS \cite{The Higgs boson spin and CP parity in the decays h to ZZ or WW or gamma gamma and constraints on the hVV couplings ATLAS} collaborations. For this purpose we first express our $XZZ$ couplings in terms of the CMS ones $\tilde{a}_1$, $\tilde{a}_2$, $\tilde{a}_3$ (we denote $a_1, a_2, a_3$ from \cite {Probabilities of Higgs boson spin-parity hypotheses and constraints on the hVV couplings CMS} as $\tilde{a}_1$, $\tilde{a}_2$, $\tilde{a}_3$ to avoid confusion):
\begin{subequations}
\label{a_Z, b_Z, c_Z in terms of tilde a_1, tilde a_2, tilde a_3, Lambda_1, phi_Lambda_1}
\begin{align}
\label{a_Z in terms of tilde a_1, tilde a_2, Lambda_1, phi_Lambda_1}
& a_Z = \tilde{\alpha} \left (\tilde{a}_1 - \exp (i \phi_{\Lambda_1}) \frac{a_1 + a_2}{\Lambda_1^2} + \frac{m_X^2 - a_1 - a_2}{m_Z^2} \tilde{a}_2 \right), \\
\label{b_Z in terms of tilde a_2}
& b_Z = -2 \tilde{\alpha} \frac{m_X^2}{m_Z^2} \tilde{a}_2, \\
\label{c_Z in terms of tilde a_3}
& c_Z = -i \tilde{\alpha} \frac{m_X^2}{m_Z^2} \tilde{a}_3,
\end{align}
\end{subequations}
where $\tilde{\alpha} \equiv \alpha_0 {v}/{2}$, $\alpha_0$ is the proportionality factor of the amplitude $A (HZZ)$ of the transition $X \to Z_1^* Z_2^*$ (see Eq.~(1) in \cite{Probabilities of Higgs boson spin-parity hypotheses and constraints on the hVV couplings CMS}), $v \equiv {1}/{\sqrt{\sqrt{2} G_F}}$ is the vacuum expectation value of the Higgs field, $\Lambda_1$ is a scale of physics beyond the SM, $\phi_{\Lambda_1}$ is the phase in the term with $\Lambda_1$. In general $\tilde{a}_1$, $\tilde{a}_2$, $\tilde{a}_3$ may depend on $a_1$ and $a_2$, however in \cite {Probabilities of Higgs boson spin-parity hypotheses and constraints on the hVV couplings CMS} they are set to be constant. The ATLAS $XZZ$ couplings $\alpha$, $\kappa_{SM}$, $\kappa_{HZZ}$, $\kappa_{AZZ}$ are related to the CMS ones in the following way:
\begin{align}
\label{a relation between the ATLAS XZZ couplings and the CMS ones}
\tilde{\alpha} \left (\tilde{a}_1 - \exp (i \phi_{\Lambda_1}) \frac{a_1 + a_2}{\Lambda_1^2} \right) = \kappa_{SM} \cos \alpha, ~~ \tilde{\alpha} \tilde{a}_2 = \frac{v}{4 \Lambda} \kappa_{HZZ} \cos \alpha, ~~ \tilde{\alpha} \tilde{a}_3 = \frac{v}{4 \Lambda} \kappa_{AZZ} \sin \alpha,
\end{align}
where $\Lambda$ is the EFT energy scale. Note that comparing the Lagrangian (1) in \cite{The Higgs boson spin and CP parity in the decays h to ZZ or WW or gamma gamma and constraints on the hVV couplings ATLAS} with the one describing the interaction of the SM Higgs field with $ZZ$ and $W^- W^+$, one can deduce that the coupling $g_{HZZ} $ from (1) in \cite{The Higgs boson spin and CP parity in the decays h to ZZ or WW or gamma gamma and constraints on the hVV couplings ATLAS} is equal to $2 m_Z^2 / v$. In \cite{The Higgs boson spin and CP parity in the decays h to ZZ or WW or gamma gamma and constraints on the hVV couplings ATLAS} the couplings $\alpha$, $\kappa_{SM}$, $\kappa_{HZZ}$, $\kappa_{AZZ}$ are considered constant and real.

In Refs.~\cite {Probabilities of Higgs boson spin-parity hypotheses and constraints on the hVV couplings CMS, The Higgs boson spin and CP parity in the decays h to ZZ or WW or gamma gamma and constraints on the hVV couplings ATLAS} 95\% CL allowed regions for $hZZ$ couplings are reported (see Table~\ref{CMS and ATLAS 95 percent CL allowed regions for hZZ couplings}). Note that
\begin{align}
& \tilde{\kappa}_{HZZ} \equiv \frac{v}{4 \Lambda} \kappa_{HZZ}, \qquad \tilde{\kappa}_{AZZ} \equiv \frac{v}{4 \Lambda} \kappa_{AZZ}, & \\
& \frac{\tilde{\kappa}_{HZZ}} {\kappa_{SM}} = - \frac{m_Z^2 \, b_Z}{2m_X^2 \, a_Z + (m_X^2 - a_1 - a_2) \, b_Z}, \qquad \frac{\tilde{\kappa}_{AZZ}} {\kappa_{SM}} \tan \alpha = \frac{2 i m_Z^2 \, c_Z}{2 m_X^2 \, a_Z + (m_X^2 - a_1 - a_2) \, b_Z} &
\end{align}
and in the limit $\Lambda_1 \to \infty$ the CMS and ATLAS ratios coincide:
\begin{align}
\frac{\tilde{a}_2}{\tilde{a}_1} = \lim_{\Lambda_1 \to \infty} \frac{\tilde{\kappa}_{HZZ}} {\kappa_{SM}}, \qquad \frac{\tilde{a}_3}{\tilde{a}_1} = \lim_{\Lambda_1 \to \infty} \left (\frac{\tilde{\kappa}_{AZZ}} {\kappa_{SM}} \tan \alpha \right).
\end{align}
%

\begin{table}[h]
\caption{The CMS \cite {Probabilities of Higgs boson spin-parity hypotheses and constraints on the hVV couplings CMS} and ATLAS \cite{The Higgs boson spin and CP parity in the decays h to ZZ or WW or gamma gamma and constraints on the hVV couplings ATLAS} 95\% CL allowed regions for $hZZ$ couplings. The last row shows the conditions under which these regions have been derived.}
\label{CMS and ATLAS 95 percent CL allowed regions for hZZ couplings}
\begin{tabular}{c c |  c c}
\hline \hline
\multicolumn{2}{c|}{CMS} & \multicolumn{2}{c}{ATLAS} \\
\hline
$\frac{\tilde{a}_2}{\tilde{a}_1}$ & $\frac{\tilde{a}_3}{\tilde{a}_1}$ & $\frac{\tilde{\kappa}_{HZZ}} {\kappa_{SM}}$ & $\frac{\tilde{\kappa}_{AZZ}} {\kappa_{SM}} \tan \alpha$ \\
\hline
$[-2.28, -1.88] \cup [-0.69, \infty)$ & [-2.05, 2.19] & (-0.75, 2.45) & (-2.85, 0.95) \\
${\rm Im} \frac{\tilde{a}_2}{\tilde{a}_1} = 0$, $\phi_{\Lambda_1} = 0$ or $\pi$ & ${\rm Im} \frac{\tilde{a}_3}{\tilde{a}_1} = 0$, $\phi_{\Lambda_1} = 0$ or $\pi$ & $\kappa_{AZZ} = 0$ & $\kappa_{HZZ} = 0$ \\
\hline \hline
\end{tabular}
\end{table}

Following \cite{The Higgs boson spin and CP parity in the decays h to ZZ or WW or gamma gamma and constraints on the hVV couplings ATLAS}, we assume the ATLAS $hZZ$ couplings to be constant. Then considering the case $\kappa_{HZZ} = 0$, we find that our couplings $a_{hZ}$, $b_{hZ}$, $c_{hZ}$ are constant as well (see (\ref{a_Z, b_Z, c_Z in terms of tilde a_1, tilde a_2, tilde a_3, Lambda_1, phi_Lambda_1}), (\ref{a relation between the ATLAS XZZ couplings and the CMS ones})), and using (\ref{an allowed interval of c at |a| = 1 and b = 0}) we obtain an allowed interval for $\tilde{\kappa}_{AZZ} \tan \alpha / \kappa_{SM}$ (see Table~\ref{our allowed regions for hZZ couplings}). However, in case $\kappa_{HZZ} = 0$ the results (\ref{an allowed interval of b at a = 1 and c = 0 and Im b = 0}) and (\ref{an allowed interval of Im b at a = 1 and c = 0 and Re b = 0}) only show that $h$ may be the SM Higgs boson, and thus they do not constrain any $hZZ$ couplings.

If $\kappa_{HZZ} \neq 0$, then $a_{hZ}$ acquires a dependence on the invariant masses squared $a_1$ and $a_2$, and therefore the constraints (\ref{a constraint on a, b, c according to experimental data of the CMS}) and (\ref{possible sets of values of a, b, c}) get invalid since they have been derived under the assumption that $|a_{hZ}^{\prime}|$ is independent of $a_2$. Therefore to constrain the ATLAS couplings in case $\kappa_{HZZ} \neq 0$, we start with Eqs.~(\ref{(sigma by the branching) over (sigma in the SM by the branching in the SM)}) and (\ref{sigma (pp->h) by Gamma_h approx sigma_SM (pp->h) by Gamma_h SM}), which demonstrate that within one standard deviation
\[ \Gamma (h \rightarrow Z_1^* Z_2^* \rightarrow 4 l) / \Gamma_{SM} (h \rightarrow Z_1^* Z_2^* \rightarrow 4 l) \in [0.68, \, 1.22]. \]
To obtain $\Gamma$ we have to calculate the integral (\ref{approximate formulas for Gamma}) for $a_Z^{\prime}$ depending on $a_2$. Taking into account the limits of the integration, we substitute $a_2$ with $(m_X - m_Z)^2 / 2$ in the expression for $a_Z^{\prime}$ (see (\ref{a_Z in terms of tilde a_1, tilde a_2, Lambda_1, phi_Lambda_1}), (\ref{a relation between the ATLAS XZZ couplings and the CMS ones})) and therefore derive Eq.~(\ref{a formula for Gamma when a_Z, b_Z, c_Z are constants and f_1 and f_2 are arbitrary}) where $a_Z$ has the expression (\ref{a_Z in terms of tilde a_1, tilde a_2, Lambda_1, phi_Lambda_1}) with $a_2 = (m_X - m_Z)^2 / 2$. It means that if $\kappa_{HZZ}$ is not zero, we may use (\ref{a constraint on a, b, c according to experimental data of the CMS}) and (\ref{an allowed interval of b at a = 1 and c = 0 and Im b = 0}), (\ref{an allowed interval of Im b at a = 1 and c = 0 and Re b = 0}) with $a_{hZ}^{\prime}$ determined by Eq.~(\ref{a_Z in terms of tilde a_1, tilde a_2, Lambda_1, phi_Lambda_1}) where $a_2$ is replaced by $(m_h - m_Z)^2 / 2$. This conclusion allows us to constrain $\tilde{\kappa}_{HZZ} / \kappa_{SM}$ and ${\rm Im} \, \tilde{\kappa}_{HZZ} / {\rm Re} \, \kappa_{SM}$, as one can see in Table~\ref{our allowed regions for hZZ couplings}.

\begin{table}[h]
\caption{Our allowed regions for the ATLAS $hZZ$ couplings. The last two rows show the conditions under which these regions have been derived.}
\label{our allowed regions for hZZ couplings}
\begin{tabular}{c|c|c}
\hline \hline
$\frac{\tilde{\kappa}_{HZZ}} {\kappa_{SM}}$ & $\frac{\tilde{\kappa}_{AZZ}} {\kappa_{SM}} \tan \alpha$ & $\frac{{\rm Im} \, \tilde{\kappa}_{HZZ}} {{\rm Re} \, \kappa_{SM}}$ \\
\hline
$[-2.38, -1.89] \cup [-0.24, 1.13]$ & [-1.28, 1.28] & [-1.01, 1.01]\\
$a_2 = (m_h - m_Z)^2 / 2$ in (\ref{a_Z in terms of tilde a_1, tilde a_2, Lambda_1, phi_Lambda_1}), & $|a_{hZ}^{\prime}| = 1$, $\kappa_{HZZ} = 0$, ${\rm Im} \frac{\tilde{\kappa}_{AZZ}} {\kappa_{SM}} = 0$ & $a_2 = (m_h - m_Z)^2 / 2$ in (\ref{a_Z in terms of tilde a_1, tilde a_2, Lambda_1, phi_Lambda_1}), \\
$a_{hZ}^{\prime} = 1$, $\kappa_{AZZ} \sin \alpha = 0$, ${\rm Im} \, \kappa_{HZZ} = 0$ & & $a_{hZ}^{\prime} = 1$, $\kappa_{AZZ} \sin \alpha = 0$, ${\rm Re} \, \kappa_{HZZ} = 0$ \\
\hline \hline
\end{tabular}
\end{table}

Note that the results (\ref{a constraint on a, b, c according to experimental data of the CMS}), (\ref{possible sets of values of a, b, c}) along with the regions shown in Table~\ref{our allowed regions for hZZ couplings} are estimated with consideration of the one sigma interval in (\ref{(sigma by the branching) over (sigma in the SM by the branching in the SM)}), with the approximation (\ref{sigma (pp->h) by Gamma_h approx sigma_SM (pp->h) by Gamma_h SM}), the central values of $m_h$, $m_Z$, $\Gamma_Z$ from Table~\ref{experimental values of 'constants' related to decays X-> Z_1^* Z_2^* -> f_1 antif_1 f_2 antif_2} and Eq.~(\ref{a formula for Gamma when a_Z, b_Z, c_Z are constants and f_1 and f_2 are arbitrary}). Comparing Tables~\ref{CMS and ATLAS 95 percent CL allowed regions for hZZ couplings} and \ref{our allowed regions for hZZ couplings}, one notices significant overlaps between the constraints reported in papers \cite{The Higgs boson spin and CP parity in the decays h to ZZ or WW or gamma gamma and constraints on the hVV couplings ATLAS, Probabilities of Higgs boson spin-parity hypotheses and constraints on the hVV couplings CMS} and our ones. In addition, we present an allowed interval for the ratio ${\rm Im} \, \tilde{\kappa}_{HZZ} / {\rm Re} \, \kappa_{SM}$ unconstrained in Refs.~\cite{The Higgs boson spin and CP parity in the decays h to ZZ or WW or gamma gamma and constraints on the hVV couplings ATLAS, Probabilities of Higgs boson spin-parity hypotheses and constraints on the hVV couplings CMS}.

We choose the following sets of values of $a_{hZ}^{\prime}$, $b_{hZ}^{\prime}$ and $c_{hZ}^{\prime}$:
\begin{align}
\label{four sets of possible values of a, b, c}
& |a_{hZ}^{\prime}| = 1, \; b_{hZ}^{\prime} = 0, \; c_{hZ}^{\prime} = 0, & \notag \\
& a_{hZ}^{\prime} = 1, \; b_{hZ}^{\prime} = 0, \; c_{hZ}^{\prime} = 0.5, & \notag \\
& a_{hZ}^{\prime} = 1, \; b_{hZ}^{\prime} = 0, \; c_{hZ}^{\prime} = 0.5 i, & \notag \\
& a_{hZ}^{\prime} = 1, \; b_{hZ}^{\prime} = -0.5, \; c_{hZ}^{\prime} = 0 &
\end{align}
and
\begin{align}
\label{a = 1, b = -0,5 i, c = 0}
a_{hZ}^{\prime} = 1, \; b_{hZ}^{\prime} = -0.5 i, \; c_{hZ}^{\prime} = 0,  &
\end{align}
which are consistent with the constraints (\ref{possible sets of values of a, b, c}). The sets (\ref{four sets of possible values of a, b, c}) and (\ref{a = 1, b = -0,5 i, c = 0}) will be used for examination of further results.

Regarding the selected values in (\ref{four sets of possible values of a, b, c}) and (\ref{a = 1, b = -0,5 i, c = 0}) one should mention that even in the SM the couplings $b_{hZ}$ and $c_{hZ}$ acquire small values due to electroweak radiative corrections where ${\rm Im} \, b_{hZ}$ and ${\rm Im} \, c_{hZ}$ come from the absorptive parts of the corresponding loop diagrams. In Eqs.~(\ref{four sets of possible values of a, b, c}),  (\ref{a = 1, b = -0,5 i, c = 0}) we assume that the $hZZ$ vertex may be significantly modified by physics beyond the SM.

It is of interest to study the distribution $\frac{1}{\Gamma} \frac{d \Gamma}{da_2}$ as a function of $\sqrt{a_2}$ for various sets of $a_Z^{\prime}$, $b_Z^{\prime}$, $c_Z^{\prime}$. Here $a_Z^{\prime} \equiv a_Z (m_Z^2, a_2), b_Z^{\prime} \equiv b_Z (m_Z^2, a_2), c_Z^{\prime} \equiv c_Z (m_Z^2, a_2)$. In accordance with (\ref{the differential width with respect to a_1, a_2}), the function $\frac{1}{\Gamma} \frac{d \Gamma}{da_2}$ is independent of the final fermion state. Figure \ref{plots of the central normalized a2-width for three sets of values of a, b, c} shows this observable in case $X = h$.

\begin{figure}
\includegraphics{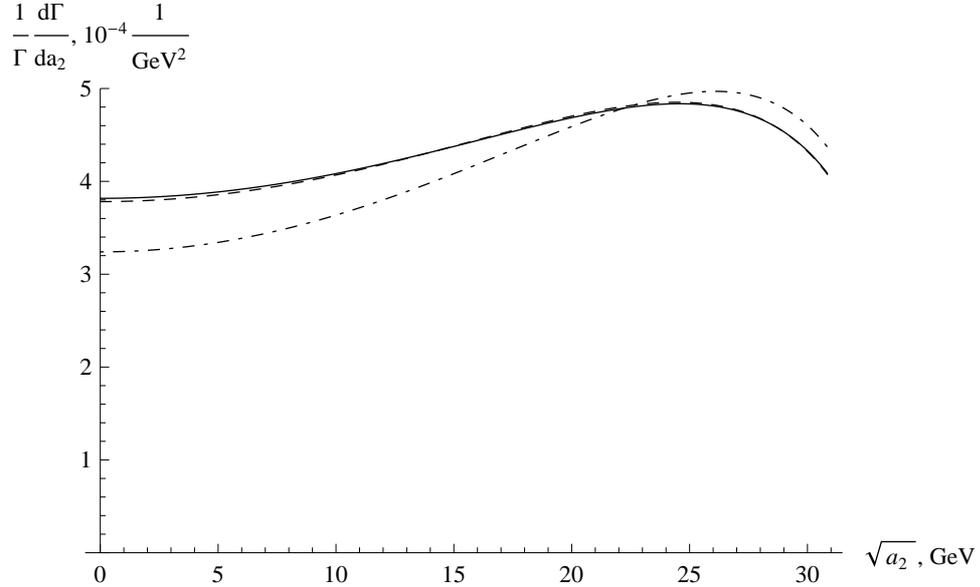}
\caption{The distribution $\frac{1}{\Gamma} \frac{d \Gamma}{da_2}$ as a function of $\sqrt{a_2}$ for the decay $h \rightarrow Z_1^* Z_2^* \rightarrow f_1 \bar{f}_1  f_2 \bar{f}_2$ in case $|a_{hZ}^{\prime}| = 1, b_{hZ}^{\prime} = 0, c_{hZ}^{\prime} = 0$ (solid line); $|a_{hZ}^{\prime}| = 1, b_{hZ}^{\prime} = 0, |c_{hZ}^{\prime}| = 0.5$ (dashed line); $a_{hZ}^{\prime} = 1, b_{hZ}^{\prime} = -0.5, c_{hZ}^{\prime} = 0$ (dash-dotted line).}
\label{plots of the central normalized a2-width for three sets of values of a, b, c}
\end{figure}

As one can see from Fig.~\ref{plots of the central normalized a2-width for three sets of values of a, b, c}, the function $\frac{1}{\Gamma} \frac{d \Gamma}{da_2}$ is sensitive to $b_{hZ}^{\prime}$ and almost insensitive to $c_{hZ}^{\prime}$. For this reason, having measured this distribution with sufficient accuracy, one can get significant constraints on the values of $b_{hZ}^{\prime}$. However, one should keep in mind that this conclusion is obtained for the case in which $|a_{hZ}^{\prime}|$, $|b_{hZ}^{\prime}|$, $|c_{hZ}^{\prime}|$ and $\cos (\arg b_{hZ}^{\prime} - \arg a_{hZ}^{\prime})$ are independent of $a_2$, and their $a_2$-dependence can considerably modify the dependence of $\frac{1}{\Gamma} \frac{d \Gamma}{da_2}$. In Sec. \ref{subsection: connection between the helicity coefficients and observables} we develop methods of getting constraints on the dependences of $a_Z^{\prime}$, $b_Z^{\prime}$, $c_Z^{\prime}$ on $a_2$.


\subsection{Connection between the helicity coefficients of the decay $X \rightarrow Z_1^* Z_2^*$ and observables}
\label{subsection: connection between the helicity coefficients and observables}

Let us consider now arbitrary dependences of $a_Z (a_1, a_2)$, $b_Z (a_1, a_2)$, $c_Z (a_1, a_2)$ such that the differential width $\frac {d^5 \Gamma} {da_1 da_2 d \theta_1 d \theta_2 d \varphi}$ has a sharp maximum as a function of $a_1$ and $a_2$ at $\sqrt{a_1} = m_Z$ or $\sqrt{a_2} = m_Z$ for any $f_1$, $f_2$.
From Eq.~(\ref{the differential width with respect to a_1, a_2, theta_1, theta_2, varphi}), using approximations analogous to those used when deriving the formulas (\ref{approximate formulas for the differential width with respect to a_2}), (\ref{approximate formulas for Gamma}), we carry out integration over $a_1$ and some of the angular variables. Then we obtain the following relations between observables $O_i (a_2)$ and the helicity coefficients:
\begin{align}
& O_1^{(1)} (a_2) \equiv  \Big(\frac{d \Gamma} {da_2} \Bigr)^{-1} \, \Bigl( \int \limits_{0}^{\frac{\pi}{2}} d \theta_1  \frac{d^2 \Gamma} {da_2 d \theta_1} - \int \limits_{\frac{\pi}{2}}^{\pi}  d \theta_1 \frac{d^2 \Gamma} {da_2 d \theta_1} \Bigr)   = - \frac{3}{2} A_{f_1} \frac{{\rm  Re}(A_{\parallel}^{\prime *} A_{\perp}^{\prime})} {\sum_{p} |A_p^{\prime}|^2},
\notag \\
& O_1^{(2)}(a_2) \equiv \Big(\frac{d \Gamma} {da_2}\Bigr)^{-1} \, \Bigl( \int \limits_{0}^{\frac{\pi}{2}} d \theta_2 \frac{d^2 \Gamma} {da_2 d \theta_2} - \int \limits_{\frac{\pi}{2}}^{\pi} d \theta_2  \frac{d^2 \Gamma} {da_2 d \theta_2} \Bigr)  = - \frac{3}{2} A_{f_2} \frac{{\rm  Re}(A_{\parallel}^{\prime *} A_{\perp}^{\prime})} {\sum_{p} |A_p^{\prime}|^2},
&
\end{align}
under the condition $\sqrt{a_2} \in \left (0, m_X - \sqrt{m_Z^2 + \varepsilon_2} \right]$.

One can write these two formulas in the following way:
\begin{align}
\label{definitions and expressions for O_1 1,2}
O_1^{(1,2)}(a_2) \equiv \Bigl( \frac{d \Gamma} {da_2} \Bigr)^{-1} \Bigl( {\int \limits_{0}^{\frac{\pi}{2}} d \theta_{1,2} \frac{d^2 \Gamma} {da_2 d \theta_{1,2}} - \int \limits_{\frac{\pi}{2}}^{\pi} d \theta_{1,2} \frac{d^2 \Gamma} {da_2 d \theta_{1,2}}} \Bigr)  = - \frac{3}{2} A_{f_{1,2}} \frac{{\rm  Re}(A_{\parallel}^{\prime *} A_{\perp}^{\prime})} {\sum_{p} |A_p^{\prime}|^2}.
\end{align}
Then we deduce that
\begin{eqnarray}
\label{a definition and an expression for O_2}
O_2 (a_2) & \equiv &
\Big( \frac{d \Gamma} {d a_2} \Bigr)^{-1} \Bigl( {\int \limits_{\frac{\pi}{2} - \beta}^{\frac{\pi}{2} - \alpha} d \theta_2 \frac{d^2 \Gamma} {d a_2 d \theta_2} + \int \limits_{\frac{\pi}{2} + \alpha}^{\frac{\pi}{2} + \beta} d \theta_2  ...   } \Bigr)
 =  \Bigl( \frac{d \Gamma} {d a_2} \Bigr)^{-1} \Bigl( {\int \limits_{\frac{\pi}{2} - \beta}^{\frac{\pi}{2} - \alpha} d \theta_1 \frac{d^2 \Gamma} {d a_2 d \theta_1} + \int \limits_{\frac{\pi}{2} + \alpha}^{\frac{\pi}{2} + \beta} d \theta_1  ...  } \Bigr)    \notag \\
& = &  \frac{1}{4} \Bigl( (\sin \beta - \sin \alpha) (3 + \sin^2 \alpha + \sin^2 \beta + \sin \alpha \sin \beta)  \notag \\
& + & 3 \frac{|A_0^{\prime}|^2}{\sum_{p} |A_p^{\prime}|^2}
(\sin \beta \cos^2 \beta - \sin \alpha \cos^2 \alpha) \Bigr), \\
& & 0 \leq \alpha < \beta \leq \frac{\pi}{2},
\notag
\end{eqnarray}
\begin{align}
\label{a definition and an expression for O_3}
O_3 (a_2) & \equiv  \Bigl( \frac{d \Gamma} {da_2} \Bigr)^{-1} \Bigl( \int \limits_{0}^{\frac{\pi}{2}} d \theta_2 \Bigl( \int \limits_{0}^{\frac{\pi}{2}} d \theta_1 \frac{d^3 \Gamma} {da_2 d \theta_1 d \theta_2} - \int \limits_{\frac{\pi}{2}}^{\pi} d \theta_1 \frac{d^3 \Gamma} {da_2 d \theta_1 d \theta_2} \Bigr) - \int \limits_{\frac{\pi}{2}}^{\pi} d \theta_2 ... \Bigr) \notag \\
& = \frac{9}{16} A_{f_1} A_{f_2} \frac{|A_{\parallel}^{\prime}|^2 + |A_{\perp}^{\prime}|^2}{\sum_{p} |A_p^{\prime}|^2},
\end{align}
\begin{align}
\label{a definition and an expression for O_4}
O_4 (a_2) & \equiv \Bigl( \frac{d \Gamma} {da_2} \Bigr)^{-1} \Bigl( {\int \limits_{0}^{\frac{\pi}{4}} d \varphi \frac{d^2 \Gamma} {da_2 d \varphi} - \int \limits_{\frac{\pi}{4}}^{\frac{3}{4} \pi} d \varphi ... + \int \limits_{\frac{3}{4} \pi}^{\frac{5}{4} \pi} d \varphi ...  - \int \limits_{\frac{5}{4} \pi}^{\frac{7}{4} \pi} d \varphi ... + \int \limits_{\frac{7}{4} \pi}^{2 \pi} d \varphi ... } \Bigr) \notag \\
& = \frac{1}{2 \pi} \frac{|A_{\parallel}^{\prime}|^2 - |A_{\perp}^{\prime}|^2}{\sum_{p} |A_p^{\prime}|^2},
\end{align}
\begin{align}
\label{a definition and an expression for O_5}
O_5 (a_2) \equiv \Bigl( \frac{d \Gamma} {da_2}\Bigr)^{-1} \Bigl( {\int \limits_{0}^{\frac{\pi}{2}} d \varphi \frac{d^2 \Gamma} {da_2 d \varphi} - \int \limits_{\frac{\pi}{2}}^{\pi} d \varphi ... + \int \limits_{\pi}^{\frac{3}{2} \pi} d \varphi ... - \int \limits_{\frac{3}{2} \pi}^{2 \pi} d \varphi ... } \Bigr) = - \frac{1}{\pi} \frac{{\rm Im}(A_{\parallel}^{\prime *} A_{\perp}^{\prime})}{\sum_{p} |A_p^{\prime}|^2},
\end{align}
\begin{align}
\label{a definition and an expression for O_6}
O_6 (a_2) \equiv \Bigl( \frac{d \Gamma} {da_2} \Bigr)^{-1} \Bigl( {\int \limits_{0}^{\frac{\pi}{2}} d \varphi \frac{d^2 \Gamma} {da_2 d \varphi} - \int \limits_{\frac{\pi}{2}}^{\frac{3}{2} \pi} d \varphi ... + \int \limits_{\frac{3}{2} \pi}^{2 \pi} d \varphi ... } \Bigr)  = \frac{9}{32} \sqrt{2} \pi A_{f_1} A_{f_2} \frac{{\rm  Re}(A_{0}^{\prime *} A_{\parallel}^{\prime})}{\sum_{p} |A_p^{\prime}|^2},
\end{align}
\begin{align}
\label{definitions and expressions for O_7 1,2}
O_7^{(1,2)} (a_2) & \equiv \Bigl(\frac{d \Gamma} {da_2}\Bigr)^{-1} \Bigl( {\int \limits_{0}^{\pi} d \varphi \Bigl (\int \limits_{0}^{\frac{\pi}{2}} d \theta_{1,2} \frac{d^3 \Gamma} {da_2 d \theta_{1,2} d \varphi} - \int \limits_{\frac{\pi}{2}}^{\pi} d \theta_{1,2} \frac{d^3 \Gamma} {da_2 d \theta_{1,2} d \varphi} \Bigr) - \int \limits_{\pi}^{2 \pi} d \varphi ... } \Bigr)  \notag \\
& = \frac{3}{8} \sqrt{2} A_{f_{2,1}} \frac{{\rm Im}(A_0^{\prime *} A_{\parallel}^{\prime})}{\sum_{p} |A_p^{\prime}|^2},
\end{align}
\begin{eqnarray}
\label{definitions and expressions for O_8 1,2}
O_8^{(1,2)} (a_2) & \equiv & \Bigl( \frac{d \Gamma} {da_2}\Bigr)^{-1} \Bigl( {\int \limits_{0}^{\frac{\pi}{2}} d \varphi \Bigl( \int \limits_{0}^{\frac{\pi}{2}} d \theta_{1,2} \frac{d^3 \Gamma} {da_2 d \theta_{1,2} d \varphi} - \int \limits_{\frac{\pi}{2}}^{\pi} d \theta_{1,2} \frac{d^3 \Gamma} {da_2 d \theta_{1,2} d \varphi} \Bigr) - \int \limits_{\frac{\pi}{2}}^{\frac{3}{2} \pi} d \varphi ... + \int \limits_{\frac{3}{2} \pi}^{2 \pi} d \varphi ... } \Bigr)  \notag \\
& = &  - \frac{3}{8} \sqrt{2} A_{f_{2,1}} \frac{{\rm  Re}(A_0^{\prime *} A_{\perp}^{\prime})}{\sum_{p} |A_p^{\prime}|^2},
\end{eqnarray}
\begin{align}
\label{a definition and an expression for O_9}
O_9 (a_2) \equiv \Bigl( {\frac{d \Gamma} {da_2}}\Bigr)^{-1} \Bigl( {\int \limits_{0}^{\pi} d \varphi \frac{d^2 \Gamma} {da_2 d \varphi} - \int \limits_{\pi}^{2 \pi} d \varphi ... } \Bigr)  = - \frac{9}{32} \sqrt{2} \pi A_{f_1} A_{f_2} \frac{{\rm  Im}(A_0^{\prime *} A_{\perp}^{\prime})}{\sum_{p} |A_p^{\prime}|^2}.
\end{align}

From the measured observables $O_i (a_2)$ one can get constraints on the dependences of the couplings $a_Z^{\prime} (a_2)$, $b_Z^{\prime} (a_2)$ and $c_Z^{\prime}(a_2)$. As for $O_2 (a_2)$, it can be measured at a fixed value of $\sqrt{a_2}$ and at various values of the parameters $\beta$ and $\alpha$. Then, after obtaining central values and uncertainties of a quantity ${|A_0^{\prime}|^2} / {\sum_p |A_p^{\prime}|^2}$ from Eq.~(\ref{a definition and an expression for O_2}) at several sets of values of $\beta$, $\alpha$, one can combine these central values and uncertainties and thereby get a value of  ${|A_0^{\prime}|^2} / {\sum_p |A_p^{\prime}|^2}$ with greater precision than in case of any particular values of $\beta$, $\alpha$.

\begin{figure}[H]
\begin{minipage}[h]{0.47 \linewidth}
\center{\includegraphics[width=0.85\linewidth]{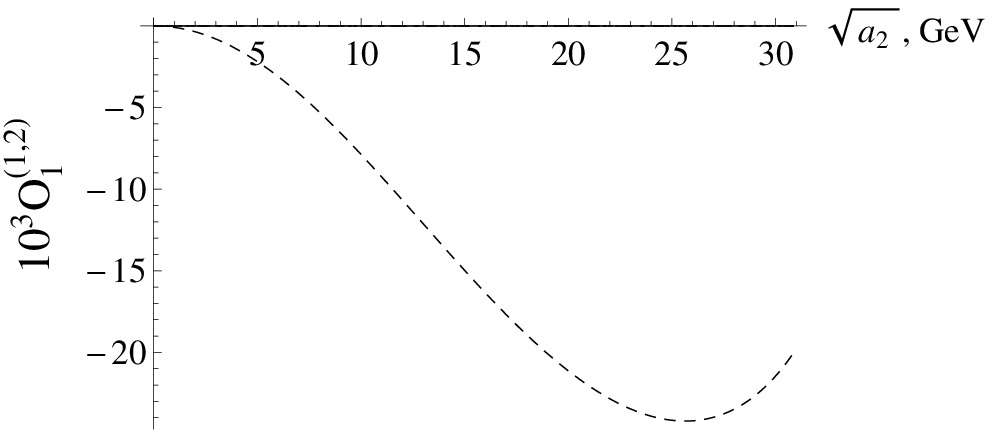}}
\end{minipage}
\hfill
\begin{minipage}[h]{0.47 \linewidth}
\center{\includegraphics[width=0.85\linewidth]{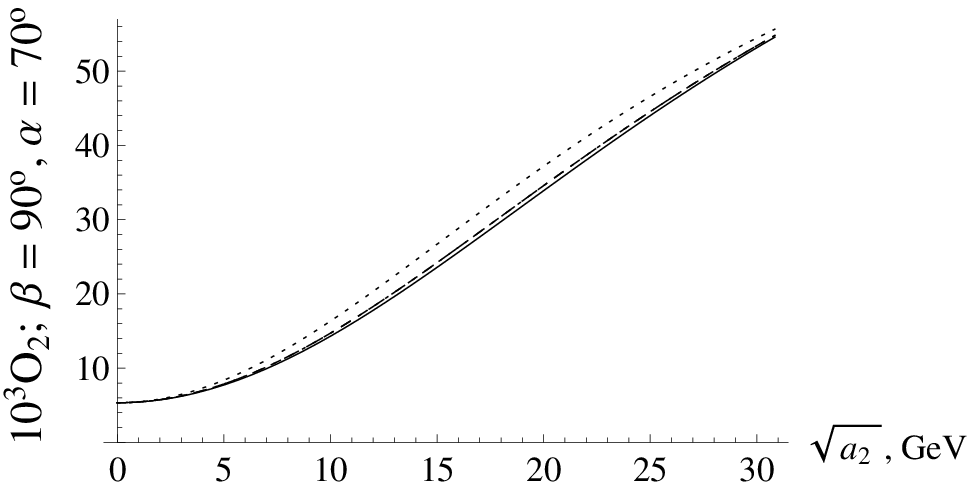}}
\end{minipage}
\vfill
\begin{minipage}[h]{0.47 \linewidth}
\center{\includegraphics[width=0.85\linewidth]{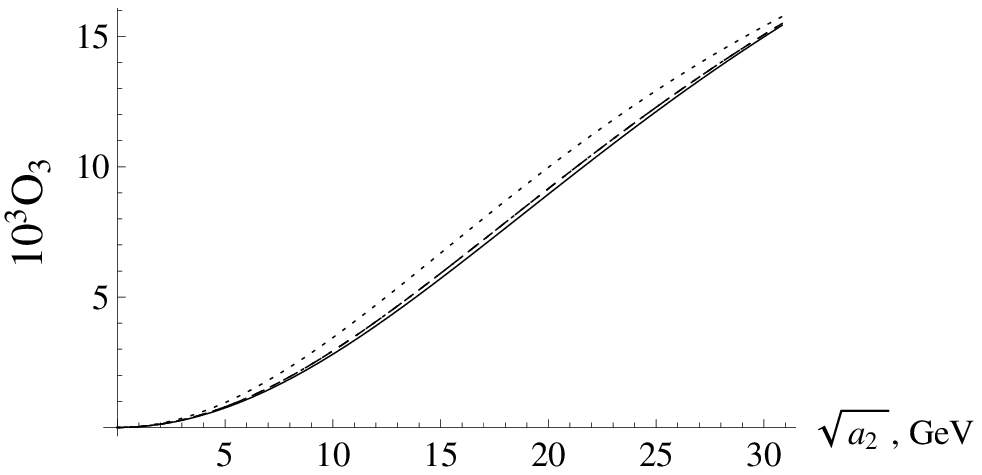}}
\end{minipage}
\hfill
\begin{minipage}[h]{0.47 \linewidth}
\center{\includegraphics[width=0.85\linewidth]{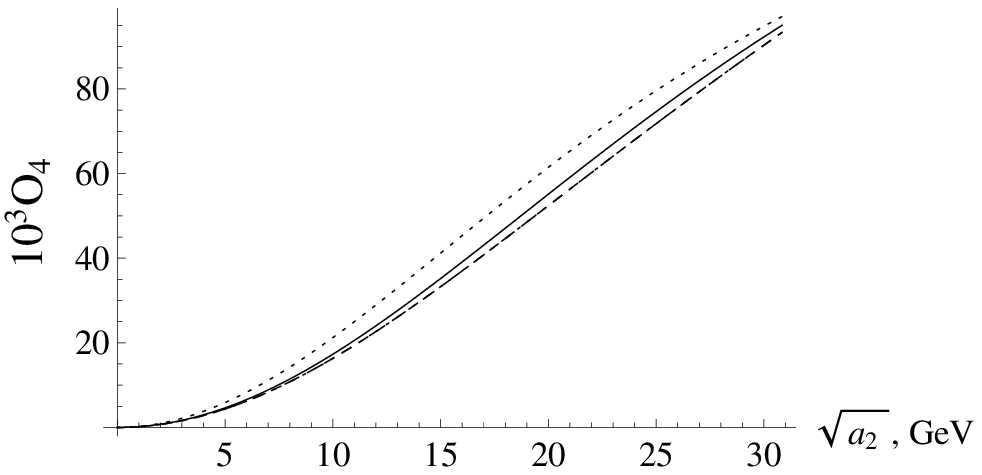}}
\end{minipage}
\vfill
\begin{minipage}[h]{0.47 \linewidth}
\center{\includegraphics[width=0.85\linewidth]{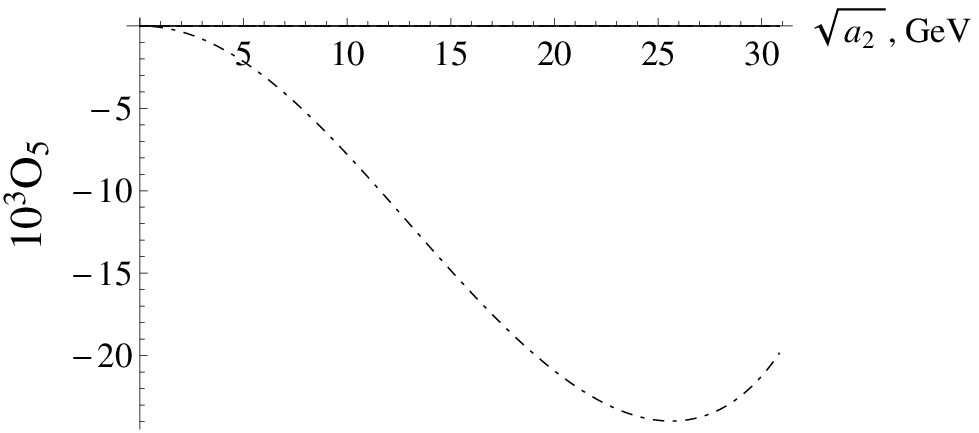}}
\end{minipage}
\hfill
\begin{minipage}[h]{0.47 \linewidth}
\center{\includegraphics[width=0.85\linewidth]{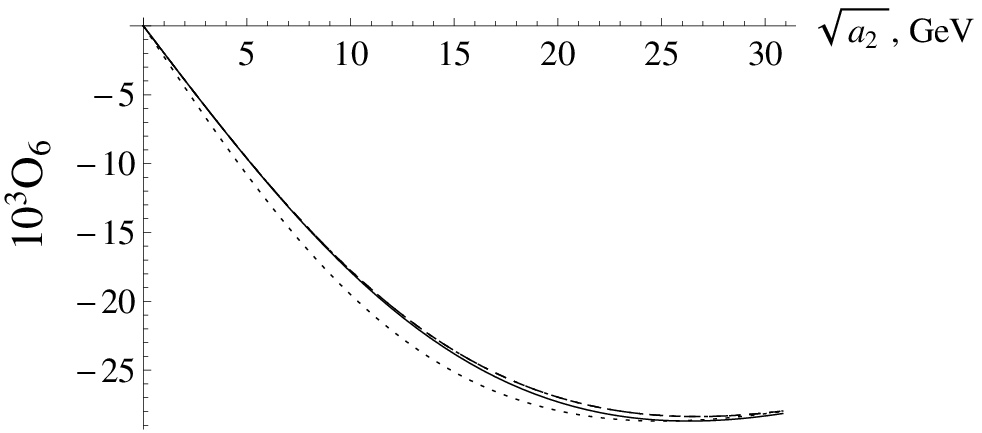}}
\end{minipage}
\vfill
\begin{minipage}[h]{0.47 \linewidth}
\center{\includegraphics[width=0.85\linewidth]{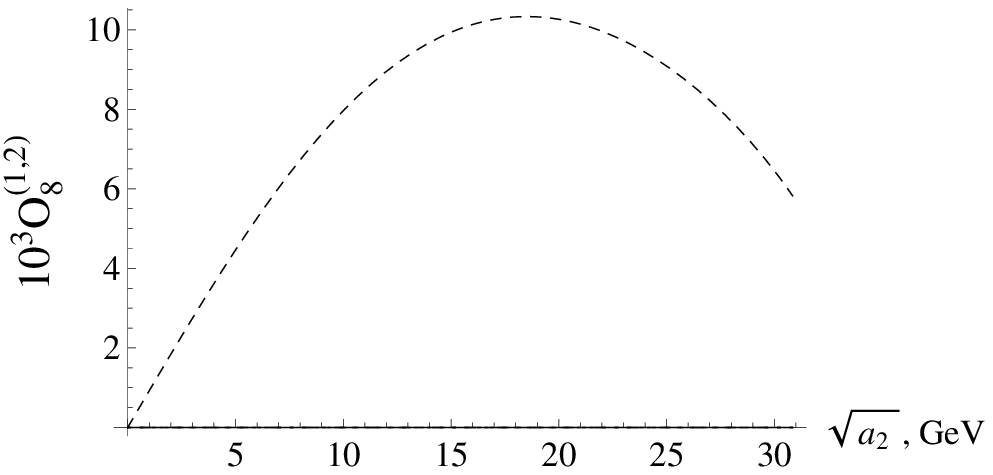}}
\end{minipage}
\hfill
\begin{minipage}[h]{0.47 \linewidth}
\center{\includegraphics[width=0.85\linewidth]{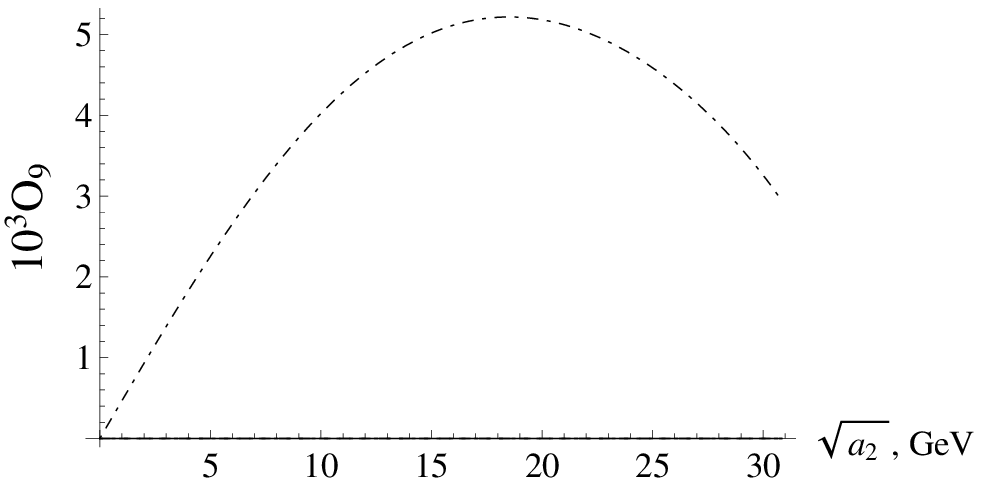}}
\end{minipage}
\caption{The observables $O_1^{(1,2)}$, $O_2$ (at $\beta = 90^{\circ}$, $\alpha = 70^{\circ}$), $O_3$, $O_4$, $O_5$, $O_6$, $O_8^{(1,2)}$, $O_9$ for the decay $h \rightarrow Z_1^* Z_2^* \rightarrow l_1^-  l_1^+  l_2^-  l_2^+ ~~ (l_j = e, \mu, \tau, l_1 \neq l_2)$ as functions of $\sqrt{a_2}$ in case $|a_{hZ}^{\prime}| = 1, b_{hZ}^{\prime} = 0, c_{hZ}^{\prime} = 0$ (solid lines); $a_{hZ}^{\prime} = 1, b_{hZ}^{\prime} = 0, c_{hZ}^{\prime} = 0.5$ (dashed lines); $a_{hZ}^{\prime} = 1, b_{hZ}^{\prime} = 0, c_{hZ}^{\prime} = 0.5 i$ (dash-dotted lines); $a_{hZ}^{\prime} = 1, b_{hZ}^{\prime} = -0.5, c_{hZ}^{\prime} = 0$ (dotted lines).}
\label{plots of all the observables except for O_7}
\end{figure}

As an illustration of the behavior of these observables, in Fig.~\ref{plots of all the observables except for O_7} we show their $\sqrt{a_2}$-dependence with the constant $a_{hZ}^{\prime}$, $b_{hZ}^{\prime}$, $c_{hZ}^{\prime}$ from the sets (\ref{four sets of possible values of a, b, c}). The observable $O_2 (a_2)$ is presented for $\beta = 90^{\circ}$ and $\alpha = 70^{\circ}$.

%
%
\begin{figure}[tbh]
\includegraphics[width=0.47\linewidth]{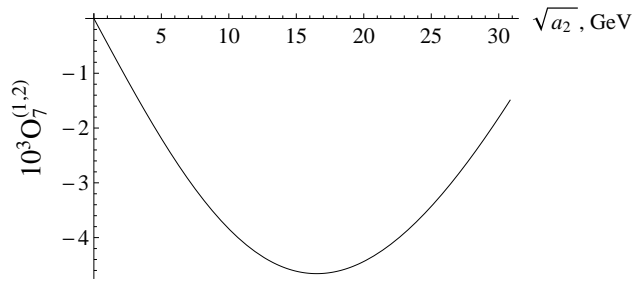}
\caption{The observables $O_7^{(1,2)}$ for the decay $h \rightarrow Z_1^* Z_2^* \rightarrow l_1^-  l_1^+  l_2^-  l_2^+ ~~ (l_j = e, \mu, \tau, l_1 \neq l_2)$ versus $\sqrt{a_2}$ in case $a_{hZ}^{\prime} = 1, b_{hZ}^{\prime} = -0.5 i, c_{hZ}^{\prime} = 0$.}
\label{a plot of O_7 for a = 1, b = -0.5 i, c = 0}
\end{figure}

As one can see, for each of the observables $O_2 (a_2)$, $O_3 (a_2)$, $O_4 (a_2)$, $O_6 (a_2)$ their dependences on $\sqrt{a_2}$ for all the four sets (\ref{four sets of possible values of a, b, c}) are very close.
The observables $O_2 (a_2) $ (at $\beta = 90^{\circ}$, $\alpha = 70^{\circ}$) and $O_4 (a_2) $ are relatively large with the maximum values greater than 0.05, and thus to measure these observables a relatively small amount of data is needed, while $O_3 (a_2)$ and $O_6(a_2)$ are smaller, which complicates their experimental observation.

Further, $O_1^{(1,2)} (a_2)$, $O_5 (a_2)$, $O_8^{(1,2)} (a_2)$, $O_9 (a_2)$ vanish for
$c_{hZ}^{\prime} (a_2) = 0$, according to Eqs.~(\ref{definitions and expressions for O_1 1,2}), (\ref{a definition and an expression for O_5}), (\ref{definitions and expressions for O_8 1,2}), (\ref{a definition and an expression for O_9}) and Fig.~\ref{plots of all the observables except for O_7}. Therefore, these observables can give significant constraints on the $CP$-odd coupling $c_{hZ}^{\prime}(a_2)$, although their moduli are relatively small.

The functions $O_7^{(1,2)} (a_2)$ are proportional to ${\rm  Im}(a_{hZ}^{\prime *} b_{hZ}^{\prime})$ (see (\ref{definitions and expressions for O_7 1,2}),  (\ref{definitions and formulae for A_0, A_parallel, A_perp for a decay X->Z_1^* Z_2^*})), and, consequently, they are equal to zero for any set from (\ref{four sets of possible values of a, b, c}).
Among all the observables under consideration, $O_7^{(1,2)} (a_2)$ are the only ones vanishing in case $b_{hZ}^{\prime}(a_2) = 0$ for any $a_{hZ}^{\prime}(a_2)$ and $c_{hZ}^{\prime}(a_2)$. Therefore, knowing the dependences $O_7^{(1,2)} (a_2)$ allows one to get notable constraints on the function $b_{hZ}^{\prime}(a_2)$. Although
in case (\ref{a = 1, b = -0,5 i, c = 0}) these observables turn out to be  relatively small in absolute value (see Fig. \ref{a plot of O_7 for a = 1, b = -0.5 i, c = 0}).

Note that from (\ref{definitions and expressions for O_1 1,2})-(\ref{a definition and an expression for O_9}), regardless of the values of the couplings $a_Z^{\prime} (a_2)$, $b_Z^{\prime} (a_2)$ and $c_Z^{\prime}(a_2)$, it follows that for any $a_2$
\begin{align}
\label{intervals of the possible values of O's}
& O_1^{(1,2)} \in [- \frac{3}{4} A_{f_{1,2}}, \, \frac{3}{4} A_{f_{1,2}}], ~~ \; O_2 \in [0, \, 1], ~~ \; O_3 \in [0, \, \frac{9}{16} A_{f_1} A_{f_2}], ~~ \; O_4, O_5 \in [- \frac{1}{2 \pi}, \, \frac{1}{2 \pi}], & \notag \\
& O_6, O_9 \in [- \frac{9}{64} \sqrt{2} \pi A_{f_1} A_{f_2}, \, \frac{9}{64} \sqrt{2} \pi A_{f_1} A_{f_2}], ~~ \; O_7^{(1,2)}, O_8^{(1,2)} \in[- \frac{3}{16} \sqrt{2} A_{f_{2,1}}, \, \frac{3}{16} \sqrt{2} A_{f_{2,1}}]. &
\end{align}

Since $A_{e^-} = A_{\mu^-} = A_{\tau^-} \approx 0.214$, $A_{\nu_e} = A_{\nu_{\mu}} = A_{\nu_{\tau}} =1$, $A_u = A_c = A_t \approx 0.697$, $A_d = A_s = A_b \approx 0.941$,
the moduli of $O_1^{(1,2)} (a_2)$, $O_7^{(1,2)} (a_2)$, $O_8^{(1,2)} (a_2)$, $O_3 (a_2)$, $O_6 (a_2)$, $O_9 (a_2)$ for the decays (\ref{X-> Z_1^* Z_2^* -> f_1 antif_1 f_2 antif_2}) with quarks and/or neutrinos in the final states are greater than those for the decays (\ref{X-> Z_1^* Z_2^* -> f_1 antif_1 f_2 antif_2}) to leptons, therefore, the former processes seem more feasible for experimental study.  On the other hand, detection of leptons is much simpler. That is why the study of each decay channel of the type (\ref{X-> Z_1^* Z_2^* -> f_1 antif_1 f_2 antif_2}) has advantages and disadvantages which strongly depend on experimental methods and parameters of detectors. Consequently, measurement of the observables $O_1^{(1,2)} (a_2)$, ..., $O_9 (a_2)$ for various decay channels and for various invariant masses of the fermion pair ($\sqrt{a_2}$) may help to put constraints on the $XZZ$ couplings $a_Z^{\prime}(a_2) $, $b_Z^{\prime}(a_2)$ and $c_Z^{\prime}(a_2)$.

\section{Conclusions}
\label{sec:Conclusions}

In the present paper the decay of a neutral particle $X$ with zero spin and arbitrary $CP$ parity into two off-mass-shell $Z$ bosons ($Z_1^*$ and $Z_2^*$) each of which decays to a fermion-antifermion pair, i.e. the decay $X \rightarrow Z_1^* Z_2^* \rightarrow f_1 \bar{f}_1 f_2 \bar{f}_2$, has been considered. The given decay has been examined at tree level for the non-identical fermions, $f_1 \neq f_2$. In the approximation of the massless fermions a formula for the fully differential width has been obtained.
It has been established that the narrow-$Z$-width approximation is applicable for finding differential decay widths of $X \rightarrow Z_1^* Z_2^* \rightarrow f_1 \bar{f}_1 f_2 \bar{f}_2$ only if the invariant mass $\sqrt{a_2}$ of the pair $f_2 \bar{f}_2$ lies in an interval $ \left(0, m_X - \sqrt{m_Z^2 + \varepsilon_2} \right]$. If the parameter $\varepsilon_2$ gets larger, the accuracy of the used approximation increases, but the interval in which the  approximation is valid reduces. As an optimal value of $\varepsilon_2$ we have chosen $\varepsilon_2 = 3 m_Z \Gamma_Z$.

In the narrow-$Z$-width approximation, but without the neglect of $\Gamma_Z$ in the propagator of $Z_2^*$, a formula for the total width of the decay (\ref{X-> Z_1^* Z_2^* -> f_1 antif_1 f_2 antif_2}) and the total width of $h \rightarrow Z_1^* Z_2^*$ have been derived. The former formula is valid in case $f_1 = f_2$ as well. Note that in Ref.~\cite{Gamma (tilde Gamma) in the SM calculated by means of the narrow-Z(W)-width approximation} within the framework of the SM the total width of the decay $X \rightarrow Z Z^* \rightarrow Z f \bar{f}$ has been found in the approximation $\Gamma_Z \approx 0$ in the propagator of $Z^*$. In an analogous way one can obtain the total width of the decay (\ref{X-> Z_1^* Z_2^* -> f_1 antif_1 f_2 antif_2}) in the SM after the neglect of $\Gamma_Z$ in the propagator of $Z_2^*$, however the formula (\ref{a formula for Gamma when a_Z, b_Z, c_Z are constants}), derived in the present paper, is more general and more precise.

Using the CMS data~\cite{h->Z_1^* Z_2^*->4l CMS}, we have found constraints on the couplings $a_{hZ}^{\prime}$, $b_{hZ}^{\prime}$, $c_{hZ}^{\prime}$, which determine the $hZZ$ interaction and the $CP$ properties of the boson $h$ detected in the experiments \cite{A discovery of a boson which looked like the SM Higgs boson}. Comparing our constraints with those reported in Refs.~\cite{Probabilities of Higgs boson spin-parity hypotheses and constraints on the hVV couplings CMS, The Higgs boson spin and CP parity in the decays h to ZZ or WW or gamma gamma and constraints on the hVV couplings ATLAS}, one can notice appreciable overlaps between the three results. Besides, we have derived an allowed interval for a ratio not studied in \cite{Probabilities of Higgs boson spin-parity hypotheses and constraints on the hVV couplings CMS, The Higgs boson spin and CP parity in the decays h to ZZ or WW or gamma gamma and constraints on the hVV couplings ATLAS}. Taking our allowed regions into account, we have selected several sets of values of the couplings $q_{hZ}^{\prime}$ ($q = a, b, c$) and analyzed results for these sets.

The observables $O_1^{(1,2)} (a_2)$, ..., $O_9 (a_2)$, measurement of which will allow one to get constraints on
the dependences of $q_Z^{\prime}$ on $\sqrt{a_2}$, are defined. It is shown that the observables $O_1^{(1,2)}(a_2)$, $O_5(a_2)$, $O_8^{(1,2)}(a_2)$, $O_9(a_2)$ become zero in case $c_Z^{\prime}(a_2) = 0$, and therefore  their experimental dependences on $\sqrt{a_2}$ can put significant constraints on the $CP$-odd coupling $c_Z^{\prime}(a_2)$. The observables $O_7^{(1,2)}(a_2)$ vanish if $b_Z^{\prime}(a_2) = 0$, and, therefore, their measurement is important for finding the $CP$-even coupling $b_Z^{\prime}(a_2)$.

Note that the absolute values of $O_1^{(1,2)}(a_2)$, $O_7^{(1,2)}(a_2)$, $O_8^{(1,2)}(a_2)$, $O_3(a_2)$, $O_6(a_2)$ and $O_9(a_2)$ for the decays (\ref{X-> Z_1^* Z_2^* -> f_1 antif_1 f_2 antif_2}) where $f_1$ and/or $f_2$ is a quark or a neutrino are greater than those for the processes in which the fermions are leptons. At the same time, the processes with the leptons are much more convenient from the experimental point of view.

Thus, measurement of the observables $O_1^{(1,2)} (a_2)$, ..., $O_9 (a_2)$ for the decays (\ref{X-> Z_1^* Z_2^* -> f_1 antif_1 f_2 antif_2}) can help to clarify the $CP$ properties of the particle $X$ and the structure of the amplitude of the decay $X \rightarrow Z_1^* Z_2^*$.

\bigskip
The authors thank Sergiy Ivashyn for useful discussions. The work is partially supported by the National Academy of Sciences of Ukraine (project ЦО-15-1/2015) and the Ministry of Education and Science of Ukraine (project 0115U00473).


\appendix

\section{Calculation of the total widths of the decays $X \rightarrow Z_1^* Z_2^* \rightarrow f_1 \bar{f}_1 f_2 \bar{f}_2$ and $h \rightarrow Z_1^* Z_2^*$}
\label{calculation of the total widths of decays X->Z_1^* Z_2^*->f_1 antif_1 f_2 antif_2 and h->Z_1^* Z_2^*}

In this Appendix we calculate the total width of the decay $X \rightarrow Z_1^* Z_2^* \rightarrow f_1 \bar{f}_1 f_2 \bar{f}_2$ for the $m_X$-values such that $m_X > m_Z $ and
for the dependences of $a_Z (a_1, a_2)$, $b_Z (a_1, a_2)$, $c_Z (a_1, a_2)$ such that the differential width $\frac {d^2 \Gamma} {da_1 da_2}$ has a sharp maximum when $\sqrt{a_1} = m_Z$ or $\sqrt{a_2} = m_Z$ and the functions $|a_Z^{\prime}|$, $|b_Z^{\prime}|$, $|c_Z^{\prime}|$, $\cos (\arg b_Z^{\prime} - \arg a_Z^{\prime})$ ($q_Z^{\prime} \equiv q_Z (m_Z^2, a_2)$; $q = a, b, c$) are independent of $a_2$. For example, $|a_Z| \approx 1$, $b_Z \approx 0$, $c_Z \approx 0$ are such dependences (see Sec.~\ref{subsec:diff_width with respect to a_1, a_2}).
Then we calculate the total decay width of $h \rightarrow Z_1^* Z_2^*$ and examine the applicability of an approximation $\Gamma_Z \approx 0$ for derivation of the total widths.


\subsection{The total width of the decay $X \rightarrow Z_1^* Z_2^* \rightarrow f_1 \bar{f}_1 f_2 \bar{f}_2$}
\label{subsec:Appendix_1}

Analogously to the derivation of Eq. (\ref{approximate formulas for the differential width with respect to a_2}), we find that
\begin{align}
\label{approximate formulas for Gamma}
\Gamma \approx \frac{2 \pi}{m_Z \Gamma_Z} \int \limits_{0}^{(m_X - m_Z)^2} da_2 f (m_Z^2, a_2).
\end{align}
Let us consider the case wherein $|a_Z^{\prime}|$, $|b_Z^{\prime}|$, $|c_Z^{\prime}|$, $\cos (\arg b_Z^{\prime} - \arg a_Z^{\prime})$ are independent of $a_2$. Having exactly calculated the integral in Eq.~(\ref{approximate formulas for Gamma}) with allowance for Eqs.~(\ref{a definition of f}) and (\ref{definitions and formulae for A_0, A_parallel, A_perp for a decay X->Z_1^* Z_2^*}), we obtain:
\begin{align}
\label{a formula for Gamma when a_Z, b_Z, c_Z are constants}
\Gamma \approx & (a_{f_1}^2 + v_{f_1}^2) (a_{f_2}^2 + v_{f_2}^2) f_0 (a_Z^{\prime}, b_Z^{\prime}, c_Z^{\prime}, m_Z, \Gamma_Z, s), &
\end{align}
where
\begin{align}
\label{a definition of f_0}
& f_0 (a_Z^{\prime}, b_Z^{\prime}, c_Z^{\prime}, m_Z, \Gamma_Z, s) \equiv  \frac{\sqrt{2} G_F^3 m_Z^7 m_X}{2^{10} 3^3 \pi^4 \Gamma_Z}  & \notag \\
& \times \Biggl [(1 - \alpha) \Bigl (-24 (23 \alpha - 5) |a_Z^{\prime}|^2 + (3 \alpha^3 - 37 \alpha^2 - \alpha (235 + 6 \beta^2) + 77 - 54 \beta^2) |b_Z^{\prime}|^2  & \notag \\
& - 16 (2 \alpha^2 + 26 \alpha - 13 + 3 \beta^2) {\rm  Re}(a_Z^{\prime *} b_Z^{\prime}) + 64 \alpha (\alpha^2 + 40 \alpha - 11 + 6 \beta^2) |c_Z^{\prime}|^2 \Bigr) + 6 \ln \left (\frac{1}{\alpha} \right)    & \notag \\
& \times \Bigl (4 (12 \alpha^2 - 18 \alpha + 3 - \beta^2) |a_Z^{\prime}|^2 + (30 \alpha^2 - 10 \alpha (3 - \beta^2) + 5 - 10 \beta^2 + \beta^4) |b_Z^{\prime}|^2  & \notag \\
& + 8 (6 \alpha^2 - \alpha (9 - \beta^2) + 2 - 2 \beta^2)
 {\rm  Re}(a_Z^{\prime *} b_Z^{\prime}) - 32 \alpha (6 \alpha^2 - \alpha (9 - \beta^2) + 1 - 3 \beta^2) |c_Z^{\prime}|^2 \Bigr)  & \notag \\
& + s \frac{3 \sqrt{2}} {\beta}\Biggl (P (\alpha, \beta, a_Z^{\prime}, b_Z^{\prime}, c_Z^{\prime}, r_+, - 4 \beta r_-)   & \notag \\
& \times \ln \frac{(1 - \alpha)^2 \sqrt{(4 \alpha - 1 + \beta^2)^2 + 4 \beta^2} + (3 \alpha - 1)^2 + \beta^2 (\alpha + 1)^2 + s \sqrt{2} (1 - \alpha) ((3 \alpha - 1) r_- - \beta (\alpha + 1) r_+)}{4 \alpha (\alpha^2 + \beta^2)}  & \notag \\
& + 2 P (\alpha, \beta, a_Z^{\prime}, b_Z^{\prime}, c_Z^{\prime}, r_-, 4 \beta r_+) \Bigl (\pi - \arg (- \alpha (3 \alpha  - 1 + \beta^2) - \beta^2 + s \frac{1 - \alpha} {\sqrt{2}} (\beta r_+ - \alpha r_-)  & \notag \\
& + i (1 - \alpha) (s \frac{\alpha r_+ + \beta r_-}{\sqrt{2}} - \beta (1 - \alpha))) \Bigr) \Biggr) \Biggr], &
\end{align}
$\alpha \equiv \left (\frac{m_Z}{m_X} \right)^2$, $\beta \equiv \frac{m_Z \Gamma_Z}{m_X^2}$,
\begin{align}
P (\alpha, \beta, a_Z^{\prime}, b_Z^{\prime}, c_Z^{\prime}, x, y) \equiv & \, 2 (2x (12 \alpha^2 - 4 \alpha + 1 - \beta^2) + y (6 \alpha - 1)) |a_Z^{\prime}|^2 + (x (16 \alpha^2 - 8 \alpha (1 - \beta^2) + 1  & \notag \\
& - 6 \beta^2 + \beta^4) + y (4 \alpha - 1 + \beta^2)) |b_Z^{\prime}|^2 + (4x (8 \alpha^2 - 2 \alpha (3 - \beta^2) + 1 - 3 \beta^2) & \notag \\
&+ y (8 \alpha - 3 + \beta^2)) {\rm  Re} (a_Z^{\prime *} b_Z^{\prime})
 - 8 \alpha (4x (4 \alpha^2 - \alpha (1 - \beta^2) - 2 \beta^2) & \notag \\
& + y (6 \alpha - 1 + \beta^2)) |c_Z^{\prime}|^2, &
\end{align}
\begin{align}
r_{\pm} \equiv \sqrt{\sqrt{(4 \alpha - 1 + \beta^2)^2 + 4 \beta^2} \pm (4 \alpha - 1 + \beta^2)}.
\end{align}
In place of $s$ one may take 1 or -1 ($f_0 (a_Z^{\prime}, b_Z^{\prime}, c_Z^{\prime}, m_Z, \Gamma_Z, -1) = f_0 (a_Z^{\prime}, b_Z^{\prime}, c_Z^{\prime}, m_Z, \Gamma_Z, 1)$). In this article the argument $\arg z$ of a complex number $z$ is defined as follows:
\begin{align}
\label{a definition of arg(z)}
& \arg z = \arctan \frac{{\rm  Im}\,z}{{\rm  Re} \,z} + \pi n ({\rm  Re} \,z,  \, {\rm  Im}\,z) ~~~ \forall z \in C | {\rm  Re} \,z \neq 0, & \notag \\
& \arg z = \pi \left(\frac{1}{2} + \Theta (- {\rm  Im}\,z) \right) ~~ \forall z \in C | ({\rm   Re} \,z =0 ~~ {\rm and} ~~ {\rm   Im}\,z \neq 0), &
\end{align}
where $n (x, y) \equiv \Theta (-x) + 2 \, \Theta (x) \Theta (-y) ~~ \forall x \neq 0$,
\begin{align}
\Theta (x) \equiv 0 ~~ \forall x \in (-\infty, 0], ~~\Theta (x) \equiv 1 ~~ \forall x \in (0, +\infty).
\end{align}
From the definition (\ref{a definition of arg(z)}) it follows that $\arg z$ is the angle counted clockwise on the
complex plane from the vector $({\rm Re} \,z, \, {\rm Im}\,z)$ towards the vector $(1, \, 0)$ and $\arg z \in [0, 2 \pi)$. Sometimes in literature a different function
\begin{align}
\label{a definition of arg^prime(z)}
\arg^{\prime} z \equiv \arg z - 2 \pi \Theta (- {\rm   Im}\,z)
\end{align}
is used as the argument of $z$. From (\ref{a definition of arg^prime(z)}) it follows that $\arg^{\prime} z \in (- \pi, \pi]$. Note that we have already used $\arg z$ above in the expression $\cos (\arg b_Z - \arg a_Z)$, but since $\cos (\arg b_Z - \arg a_Z) = \cos (\arg^{\prime} b_Z - \arg^{\prime} a_Z)$,  at that point the distinction between $\arg z$ and $\arg^{\prime} z$ was irrelevant.

Calculating the integral over $a_2$ in Eq.~(\ref{approximate formulas for Gamma}), one finds an antiderivative of $f (m_Z^2, a_2)$ on the interval $[0, (m_X - m_Z)^2]$. In this antiderivative the function $\arg u_1 (a_2)$ naturally appears, where $u_1 (a_2)$ is a complex-valued dimensionless function such that
\begin{align}
\label{properties of u_1}
& \forall a_2 \in [0, (m_X - m_Z)^2) ~~{\rm Im} \,u_1 (a_2) \neq 0, & \notag \\
& {\rm Im} \,u_1 ((m_X - m_Z)^2) = 0, ~{\rm Re} \,u_1 ((m_X - m_Z)^2) < 0. &
\end{align}
$\arg^{\prime} u_1 (a_2)$ does not emerge in place of $\arg u_1 (a_2)$ since, according to (\ref{a definition of arg^prime(z)}), the function $\arg^{\prime} z$ has a discontinuity on the half-line ${\rm Im} \, z = 0, {\rm Re} \, z < 0$ and thus $\arg^{\prime} u_1 (a_2)$ has a discontinuity at the point $a_2 = (m_X - m_Z)^2$. To avoid this drawback it is convenient to use $\arg z$ in   Eq.~(\ref{a definition of f_0}).

Note that in case of the Higgs boson, i.e. $X = h$,   Eq.~(\ref{a definition of f_0}) can also be written in terms of the function $\arg^{\prime} $: for this  one has to substitute in Eq.~(\ref{a definition of f_0})  $\pi - \arg ... $ by $s \pi - \arg^{\prime} ... $, since according to (\ref{a definition of arg^prime(z)}) and to data of Table \ref{experimental values of 'constants' related to decays X-> Z_1^* Z_2^* -> f_1 antif_1 f_2 antif_2}, $\pi - \arg ... = s \pi - \arg^{\prime} ... $.

In case of the identical fermions, $f_1 = f_2$, one may neglect the interference term and then in order to obtain a formula for $\Gamma$ one has to multiply the right-hand side of the relation (\ref{a formula for Gamma when a_Z, b_Z, c_Z are constants}) by $\frac{1}{2! 2!}$ (in view of the identity of the final fermions) and by 2 (since the contribution of the diagram with the permutation of the particles to $\Gamma$ is equal to that of the diagram without the permutation), i.e. to multiply the right-hand side by $\frac{1}{2}$. Consequently, for any $f_1$ and $f_2$
\begin{align}
\label{a formula for Gamma when a_Z, b_Z, c_Z are constants and f_1 and f_2 are arbitrary}
\Gamma \approx & (1 - \frac{1}{2} \delta_{f_1 f_2}) (a_{f_1}^2 + v_{f_1}^2) (a_{f_2}^2 + v_{f_2}^2) f_0 (a_Z^{\prime}, b_Z^{\prime}, c_Z^{\prime}, m_Z, \Gamma_Z, s) \equiv \Gamma_{\Gamma_Z}, &
\end{align}
where $\delta_{f_1 f_2} \equiv 0 \,(1)$ at $f_1 \neq f_2$ ($f_1 = f_2$).
The neglected interference term seems small based on qualitative arguments of Ref.~\cite{Romao:1998sr}. For a quantitative estimate we can use Ref.~\cite{The contribution of the interference term to Gamma (h to ZZ to 4e) in the SM at tree level} (see Table 1 there), according to which the interference contribution to $\Gamma (h \to Z_1^* Z_2^* \to 4 e)$ in the SM at tree level is  5.80\% for $m_h = 140$ GeV.

In Ref.~\cite{Gamma (tilde Gamma) in the SM calculated by means of the narrow-Z(W)-width approximation} the width of the decay $h \rightarrow Z Z^* \rightarrow Z f \bar{f}$ has been derived at tree level in the SM after the neglect of $\Gamma_Z$ in the propagator of $Z^*$. Following \cite{Gamma (tilde Gamma) in the SM calculated by means of the narrow-Z(W)-width approximation}, when calculating the integral in Eq.~(\ref{approximate formulas for Gamma}), in the expression for $f (m_Z^2, a_2)$ we may also neglect $\Gamma_Z$, and then we obtain the following approximate formula for $\Gamma$ in the SM:
\begin{align}
\label{a formula for Gamma when a_Z, b_Z, c_Z are constants in an approximation Gamma_Z = 0 in the SM}
\Gamma |_{SM} \approx & (1 - \frac{1}{2} \delta_{f_1 f_2}) \frac{\sqrt{2} G_F^3 m_Z^7 m_X}{2^7 3^2 \pi^4 \Gamma_Z} (a_{f_1}^2 + v_{f_1}^2) (a_{f_2}^2 + v_{f_2}^2)  & \notag \\
& \times \left (6 \frac{1 - 8 \alpha + 20 \alpha^2}{\sqrt{4 \alpha - 1}} \arccos \left (\frac{3 \alpha - 1}{2 \alpha^{\frac{3}{2}}} \right) - \frac{1 - \alpha}{\alpha} (2 - 13 \alpha + 47 \alpha^2) + 3 (1 - 6 \alpha + 4 \alpha^2) \ln \frac{1}{\alpha} \right) & \notag \\
& \equiv \Gamma_0 |_{SM}. &
\end{align}
From (\ref{a formula for Gamma when a_Z, b_Z, c_Z are constants in an approximation Gamma_Z = 0 in the SM}) and (\ref{a formula for Gamma when a_Z, b_Z, c_Z are constants and f_1 and f_2 are arbitrary}) we obtain that at $m_X = m_h$
\begin{align}
\label{in the SM Gamma_0 approx 1,001 Gamma_Gamma_Z}
\Gamma_0 |_{SM} \approx 1.001 \times \Gamma_{\Gamma_Z} |_{SM}.
\end{align}
Besides, $\Gamma_0 > \Gamma_{\Gamma_Z}$  (for any $ a_Z^{\prime}, b_Z^{\prime}, c_Z^{\prime}, \, f_1, f_2$) since when deriving the formula for $\Gamma_0$ one neglects the width $\Gamma_Z$ in  $f(m_Z^2, a_2)$ and the value of the integral increases. Still according to (\ref{in the SM Gamma_0 approx 1,001 Gamma_Gamma_Z}), the difference between $\Gamma_0 |_{SM}$ and $\Gamma_{\Gamma_Z} |_{SM}$ is about one per mille.

Finally, note that at $m_X = m_h$ we can represent the dependence of the function $f_0$ on the $X ZZ$ couplings $a_Z^{\prime}, b_Z^{\prime}, c_Z^{\prime}$  in the convenient form:
\begin{align}
\label{the explicit dependence of f_0 on a, b, c in case m_X = m_h}
f_0 (a_Z^{\prime}, b_Z^{\prime}, c_Z^{\prime}, m_Z, \Gamma_Z, s) \approx \left (3.359 |a_Z^{\prime}|^2 + 0.052 |b_Z^{\prime}|^2 + 0.594 \, {\rm  Re} (a_Z^{\prime *} b_Z^{\prime}) + 0.125 |c_Z^{\prime}|^2 \right) ~{\rm keV}.
\end{align}


\subsection{The total width of the decay $h \rightarrow Z_1^* Z_2^*$}
\label{subsec:total width}

The total decay width $\Gamma (h \rightarrow Z_1^* Z_2^*)$ is
\begin{align}
\label{Gamma (h->Z_1^* Z_2^*)}
\Gamma (h \rightarrow Z_1^* Z_2^*) = \sum_{f_1}  \sum_{f_2 \ge f_1} \Gamma |_{m_X = m_h},
\end{align}
where the sums run over the fermions  $e^-, \, \mu^-, \, \tau^-, \, \nu_e, \, \nu_{\mu}, \, \nu_{\tau}, \, u_i, \, c_i, \, d_i, \, s_i, \, b_i$ (since $m_h \in (4 m_b, 2 m_t]$), $i = r, g, b$ is an index of quark color. It follows from Eqs.~(\ref{Gamma (h->Z_1^* Z_2^*)}), (\ref{in the SM Gamma_0 approx 1,001 Gamma_Gamma_Z}) that in the SM
\begin{align}
\Gamma_0 (h \rightarrow Z_1^* Z_2^*) \approx 1.001 \times \Gamma_{\Gamma_Z} (h \rightarrow Z_1^* Z_2^*).
\end{align}
Further we use Eq.~(\ref{a formula for Gamma when a_Z, b_Z, c_Z are constants and f_1 and f_2 are arbitrary})  since it is more precise than Eq.~(\ref{a formula for Gamma when a_Z, b_Z, c_Z are constants in an approximation Gamma_Z = 0 in the SM}) and consider the case wherein $|a_{hZ}^{\prime}|$, $|b_{hZ}^{\prime}|$, $|c_{hZ}^{\prime}|$, $\cos (\arg b_{hZ}^{\prime} - \arg a_{hZ}^{\prime})$ do not depend on $a_2$. From Eqs.~(\ref{Gamma (h->Z_1^* Z_2^*)}), (\ref{a formula for Gamma when a_Z, b_Z, c_Z are constants and f_1 and f_2 are arbitrary}), (\ref{a definition of f_0}) we derive that 
\begin{align}
\Gamma (h \rightarrow Z_1^* Z_2^*) & \approx f_0 (a_{hZ}^{\prime}, b_{hZ}^{\prime}, c_{hZ}^{\prime}, m_Z, \Gamma_Z, s) |_{m_X = m_h} \Biggl (\frac{1}{2}\sum_{f_1 } (a_{f_1}^2 + v_{f_1}^2)^2  + & \notag \\
& + \frac{1}{2} \sum_{f_1 } \sum_{f_2 \ne f_1} (a_{f_1}^2 + v_{f_1}^2) (a_{f_2}^2 + v_{f_2}^2) \Biggr)
= \frac{f_0 (...) }{2} \left( \sum_{f} (a_f^2 + v_f^2) \right)^2 = & \notag \\
& = \frac{f_0 (...) }{18} \left (\frac{103}{2} - 100 \left (\frac{m_W}{m_Z} \right)^2 + 80 \left (\frac{m_W}{m_Z} \right)^4 \right)^2.
\end{align}
Carrying out calculations, we find the total decay width for the sets (\ref{four sets of possible values of a, b, c}), (\ref{a = 1, b = -0,5 i, c = 0}):
\begin{align}
\label{values of Gamma (h->Z_1^* Z_2^*) for various sets of values of a, b, c}
& |a_{hZ}^{\prime}| = 1, b_{hZ}^{\prime} = 0, c_{hZ}^{\prime} = 0 \Rightarrow \Gamma (h \rightarrow Z_1^* Z_2^*) \approx 91.16_{-14.50}^{+16.66} ~{\rm keV}, & \notag \\
& |a_{hZ}^{\prime}| = 1, b_{hZ}^{\prime} = 0, |c_{hZ}^{\prime}| = 0.5 \Rightarrow \Gamma (h \rightarrow Z_1^* Z_2^*) \approx 92.01_{-14.67}^{+16.85} ~{\rm keV}, & \notag \\
& a_{hZ}^{\prime} = 1, b_{hZ}^{\prime} = -0.5, c_{hZ}^{\prime}= 0 \Rightarrow \Gamma (h \rightarrow Z_1^* Z_2^*) \approx 83.45_{-13.14}^{+15.06} ~{\rm keV}, & \notag \\
& a_{hZ}^{\prime} = 1, b_{hZ}^{\prime} = \pm 0.5 i, c_{hZ}^{\prime}= 0 \Rightarrow \Gamma (h \rightarrow Z_1^* Z_2^*) \approx 91.51_{-14.57}^{+16.74} ~{\rm keV}. &
\end{align}

The uncertainties shown in Eqs.~(\ref{values of Gamma (h->Z_1^* Z_2^*) for various sets of values of a, b, c}) are calculated by finding the maximum and minimum values of the function $\Gamma (h \rightarrow Z_1^* Z_2^*)$ in the region $v \in [v_0 - 3 \sigma_v, v_0 + 3 \sigma_v]$ ($v = G_F, m_h, m_Z, m_W, \Gamma_Z$). Here $v_0$ is the central value of a quantity $v$, $\sigma_v$ is the 1-standard-deviation uncertainty of $v$; according to the data of Table~\ref{experimental values of 'constants' related to decays X-> Z_1^* Z_2^* -> f_1 antif_1 f_2 antif_2}, $G_{F0} = 1.1663787 \times 10^{-5}$~GeV$^{-2}$, $\sigma_{G_F} = 6 \times 10^{-12}$~GeV$^{-2}$, $m_{h0} = 125.7$~GeV, $\sigma_{m_h} = 0.4$~GeV etc.



\begin{thebibliography}{99}

\bibitem{A discovery of a boson which looked like the SM Higgs boson}
G. Aad et al. (ATLAS Collaboration),
Phys. Lett. B {\bf 716}, 1 (2012);
\\
S. Chatrchyan et al. (CMS Collaboration),
Phys. Lett. B {\bf 716}, 30 (2012).

\bibitem{h->Z_1^* Z_2^*->4l and h->W-* W+*->l nu l nu ATLAS}
G. Aad et al. (ATLAS Collaboration),
Phys. Lett. B {\bf 726}, 88 (2013).

\bibitem{h->Z_1^* Z_2^*->4l CMS}
S. Chatrchyan et al. (CMS Collaboration),
Phys. Rev. D {\bf 89}, 092007 (2014).

\bibitem{the probability that CPh=-1 is very small CMS}
S. Chatrchyan  et al. (CMS Collaboration),
Phys. Rev. Lett. {\bf 110}, 081803 (2013).

\bibitem{Pilaftsis:1999np}
A.~Pilaftsis, C.E.M.~Wagner,
Nucl. Phys. B {\bf 553}, 3 (1999).

\bibitem{Barger:2009pr}
V.~Barger, P.~Langacker, M.~McCaskey et al.,
Phys. Rev. D {\bf 79}, 015018 (2009).

\bibitem{Branco:2012pre}
G.C.~Branco, P.M.~Ferreira, L.~Lavoura et al.,
Phys.\ Rep.\ {\bf 516}, 1 (2012).

\bibitem{Bailin}
D.~Bailin, A.~Love, \textit{Cosmology in Gauge Field Theory and
String Theory} (Institute of Physics Publishing,
Bristol-Philadelphia, 2004).

\bibitem{Voloshin:2012}
M.B.~Voloshin,
Phys. Rev. D {\bf 86}, 093016 (2012).

\bibitem{Bishara:2013vya}
  F.~Bishara, Y.~Grossman, R.~Harnik et al.,
JHEP {\bf 1404}, 084 (2014).

\bibitem{Korchin:2013}
A.Yu.~Korchin, V.A.~Kovalchuk,
Phys. Rev. D {\bf 88}, 036009 (2013);
\\
A.Yu.~Korchin, V.A.~Kovalchuk,
Acta Phys. Polon. B {\bf 44}, 2121 (2013).

\bibitem{Gainer:2011aa}
J.~S.~Gainer, W.~Y.~Keung, I.~Low and P.~Schwaller,
Phys.\ Rev.\ D {\bf 86}, 033010 (2012).

\bibitem{Korchin:2014kha}
A.Y.~Korchin and V.A.~Kovalchuk,
Eur. Phys. J. C {\bf 74}, 3141 (2014).

\bibitem{distributions of a_2 Choi}
S.Y. Choi, D.J. Miller, M.M. M{\"u}hlleitner et al.,
Phys. Lett. B {\bf 553}, 61 (2003).

\bibitem{amplitudes of X->ZZ and X->W-W+}
V.A.~Kovalchuk,
J. Exp. Theor. Phys. {\bf 107}, 774 (2008).

\bibitem{distributions of a_2 Menon}
A. Menon, T. Modak, D. Sahoo et al.,
Phys. Rev. D {\bf 89}, 095021 (2014).

\bibitem{distributions of a_2 Sun}
Y. Sun, X.-F. Wang, and D.-N. Gao,
Int. J. Mod. Phys. A {\bf 29}, 1450086 (2014).

\bibitem{methodologies on constraining the Higgs boson couplings to ZZ W-W+ gamma gamma and Z gamma from experimental data}
A.~De Rujula, J.~Lykken, M.~Pierini et al.,
Phys. Rev. D {\bf 82}, 013003 (2010);
\\
Y.~Gao, A.~V.~Gritsan, Z.~Guo et al.,
Phys.\ Rev.\ D {\bf 81}, 075022 (2010);
\\
S.~Bolognesi, Y.~Gao, A.~V.~Gritsan et al.,
Phys. Rev. D {\bf 86}, 095031 (2012);
\\
D.~Stolarski and R.~Vega-Morales,
Phys. \ Rev. \ D {\bf 86}, 117504 (2012);
\\
P.~Avery, D.~Bourilkov, M.~Chen et al.,
Phys. Rev. D {\bf 87}, no. 5, 055006 (2013);
\\
M.~Chen, T.~Cheng, J.~S.~Gainer et al.,
Phys. Rev. D {\bf 89}, no. 3, 034002 (2014);
\\
B.~Bhattacherjee, T.~Modak, S.~K.~Patra et al.,
arXiv:1503.08924 [hep-ph].

\bibitem{the dilaton mass}
M. Gasperini,
Phys. Lett. B {\bf 327}, 214 (1994).

\bibitem{Is h the SM Higgs boson or the dilaton? number1}
Z. Chacko, R. Franceschini, and R.K. Mishra,
JHEP {\bf 1304}, 015 (2013).

\bibitem{Is h the SM Higgs boson or the dilaton? number2}
B. Bellazzini, C. Cs{\'a}ki, J. Hubisz et al.,
Eur. Phys. J. C {\bf 73}, 2333 (2013).

\bibitem{Is h the SM Higgs boson or the dilaton? (a continuation of number2)}
J. Serra,
EPJ Web Conf. {\bf 60}, 17005 (2013).

\bibitem{Gamma (tilde Gamma) in the SM calculated by means of the narrow-Z(W)-width approximation}
W.-Y. Keung and W.J. Marciano,
Phys. Rev. D {\bf 30}, 248 (1984).

\bibitem{Observables connected with the CP properties of the Higgs boson and defined with the momenta of the fermions}
R.~M.~Godbole, D.~J.~Miller and M.~M.~M{\"u}hlleitner,
JHEP {\bf 0712}, 031 (2007).

\bibitem{the helicity formalism}
W.-Y. Keung, I. Low, and J. Shu,
Phys. Rev. Lett. {\bf 101}, 091802 (2008);
\\
T.L. Trueman,
Phys. Rev. D {\bf 18}, 3423 (1978). \\
J. R. Dell’Aquila and C. A. Nelson,
Phys. Rev. D {\bf 33}, 80 (1986).

\bibitem{Particle Data Group 2014}
K.A. Olive et al. (Particle Data Group),
Chin. Phys. C {\bf 38}, 090001 (2014).

\bibitem{the narrow-width approximation is non-applicable for calculation of certain total widths}
N.N.~Achasov and V.V.~Gubin,
JETP Lett. {\bf 62}, 191 (1995)
[Pisma Zh. Eksp. Teor. Fiz. {\bf 62}, 182 (1995)].

\bibitem{the narrow-width approximation is non-applicable for calculation of certain cross sections}
D. Berdine, N. Kauer, and D. Rainwater,
Phys. Rev. Lett. {\bf 99},111601 (2007).

\bibitem{a study of the narrow-width approximation}
C.F. Uhlemann and N. Kauer,
Nucl. Phys. B {\bf 814},195 (2009).

\bibitem{The contribution of the interference term to Gamma (h to ZZ to 4e) in the SM at tree level}
A.~Bredenstein, A.~Denner, S.~Dittmaier and M.~M.~Weber,
Phys.\ Rev.\ D {\bf 74}, 013004 (2006).

\bibitem{The total and differential cross sections of the process pp to h ATLAS}
G.~Aad et al. (ATLAS Collaboration),
arXiv:1504.05833 [hep-ex].

\bibitem{LHC 2012-2014 data on SM Higgs boson branchings and total cross sections}
LHC Higgs Cross Section Working Group,\\
https://twiki.cern.ch/twiki/bin/view/LHCPhysics/CERNYellowReportPageAt8TeV.

\bibitem{Constraints on the total width of the Higgs boson CMS}
V.~Khachatryan et al. (CMS Collaboration),
Phys.\ Lett.\ B {\bf 736}, 64 (2014).

\bibitem{Properties of the Higgs boson (theory and experiment)}
S.~Heinemeyer et al. (LHC Higgs Cross Section Working Group),
arXiv:1307.1347v2 [hep-ph].

\bibitem{an approach to getting constraints on the XZZ couplings by means of finding the couplings as a point on a sphere}
J.~S.~Gainer, J.~Lykken, K.~T.~Matchev et al.,
Phys. Rev. Lett. {\bf 111}, 041801 (2013).

\bibitem{Probabilities of Higgs boson spin-parity hypotheses and constraints on the hVV couplings CMS}
V.~Khachatryan et al. (CMS Collaboration),
Phys.\ Rev.\ D {\bf 92}, no. 1, 012004 (2015).

\bibitem{The Higgs boson spin and CP parity in the decays h to ZZ or WW or gamma gamma and constraints on the hVV couplings ATLAS}
G.~Aad et al. (ATLAS Collaboration),
arXiv:1506.05669v1 [hep-ex].

\bibitem{Romao:1998sr}
J.~C.~Rom{\~a}o and S.~Andringa,
Eur.\ Phys.\ J.\ C {\bf 7}, 631 (1999).

\end{thebibliography}
\end{document}